\documentclass[final,5p,times,twocolumn,authoryear]{elsarticle}

\usepackage{amssymb}
\usepackage{lipsum}
\usepackage{amsmath}
\usepackage{array}
\usepackage{tabularx}
\usepackage{supertabular}
\usepackage{pdflscape}
\usepackage{rotating}
\usepackage{threeparttable}
\usepackage{booktabs}
\usepackage{longtable}
\usepackage{url} 
\usepackage{tikz}
\usepackage{hyperref}
\usepackage{array}
\usepackage{float} 
\usepackage{pifont}   % for \ding
\usepackage{amssymb}  % for \checkmark

\usetikzlibrary{mindmap,shadows,trees}
\usetikzlibrary{mindmap,trees} % This loads the libraries needed for mindmaps.
\journal{AI OPEN}
\usepackage{lineno}

\begin{document}
\begin{frontmatter}

%% Title, authors and addresses

%% Title
\title{Comprehensive Analysis of Transparency and Accessibility of ChatGPT, DeepSeek, and other SoTA Large Language Models}

%% Authors
\author[1]{Ranjan Sapkota\corref{cor1}}
\author[2]{Shaina Raza}
\author[3,1]{Manoj Karkee}

%% Affiliations
\affiliation[1]{organization={Biological \& Environmental Engineering}, 
            addressline={Cornell University}, 
            city={Ithaca}, 
            postcode={14850}, 
            state={NY}, 
            country={USA}}
            
\affiliation[2]{organization={Vector Institute}, 
            addressline={Ontario}, 
            city={Tronto}, 
            postcode={ W1140-108}, 
            state={ON}, 
            country={CANADA}}

\affiliation[3]{organization={Department of Biological Systems Engineering}, 
            addressline={Washington State University},  
            state={WA}, 
            country={USA}}

%% Corresponding author
\cortext[cor1]{Manoj Karkee}
\ead{rs2672@cornell.edu}

%% Abstract
\begin{abstract}
Despite increasing discussions on open-source Artificial Intelligence (AI), existing research lacks a discussion on the  transparency and accessibility of state-of-the-art (SoTA) Large Language Models (LLMs). The Open Source Initiative (OSI) has recently released its first formal definition of open-source software. This definition, when combined with standard dictionary definitions and the sparse published literature, provide an initial framework to support broader accessibility to AI models such as LLMs, but more work is essential to capture the unique dynamics of openness in AI.  In addition, concerns about open-washing, where models claim openness but lack full transparency, has been raised, which limits the reproducibility, bias mitigation, and domain adaptation of these models. In this context, our study critically analyzes SoTA LLMs from the last five years, including ChatGPT, DeepSeek, LLaMA, and others, to assess their adherence to transparency standards and the implications of partial openness. Specifically, we examine transparency and accessibility from two perspectives: open-source vs. open-weight models. Our findings reveal that while some models are labeled as open-source, this does not necessarily mean they are fully open-sourced. Even in the best cases, open-source models often do not report model training data, and code as well as key metrics, such as weight accessibility, and carbon emissions. To the best of our knowledge, this is the first study that systematically examines the transparency and accessibility of over 100 different SoTA LLMs through the dual lens of open-source and open-weight models. The findings open avenues for further research and call for responsible and sustainable AI practices to ensure greater transparency, accountability, and ethical deployment of these models.

\end{abstract}

%%Graphical abstract
%\begin{graphicalabstract}
%\includegraphics{grabs}
%\end{graphicalabstract}

%%Research highlights
%\begin{highlights}
%\item Research highlight 1
%\item Research highlight 2
%\end{highlights}

%\begin{keyword}
%% keywords here, in the form: keyword \sep keyword, up to a maximum of 6 keywords
%Open Source \sep ChatGPT \sep DeepSeek \sep LLMs \sep Transparency \sep Open Weights \sep Large Langauge Models \sep DeepSeekR1 \sep OpenAI 
%% PACS codes here, in the form: \PACS code \sep code

%% MSC codes here, in the form: \MSC code \sep code
%% or \MSC[2008] code \sep code (2000 is the default)

%\end{keyword}

\end{frontmatter}

\section{Introduction}
\label{introduction}
\begin{figure*}[ht]
\centering
\includegraphics[width=0.75 \linewidth]{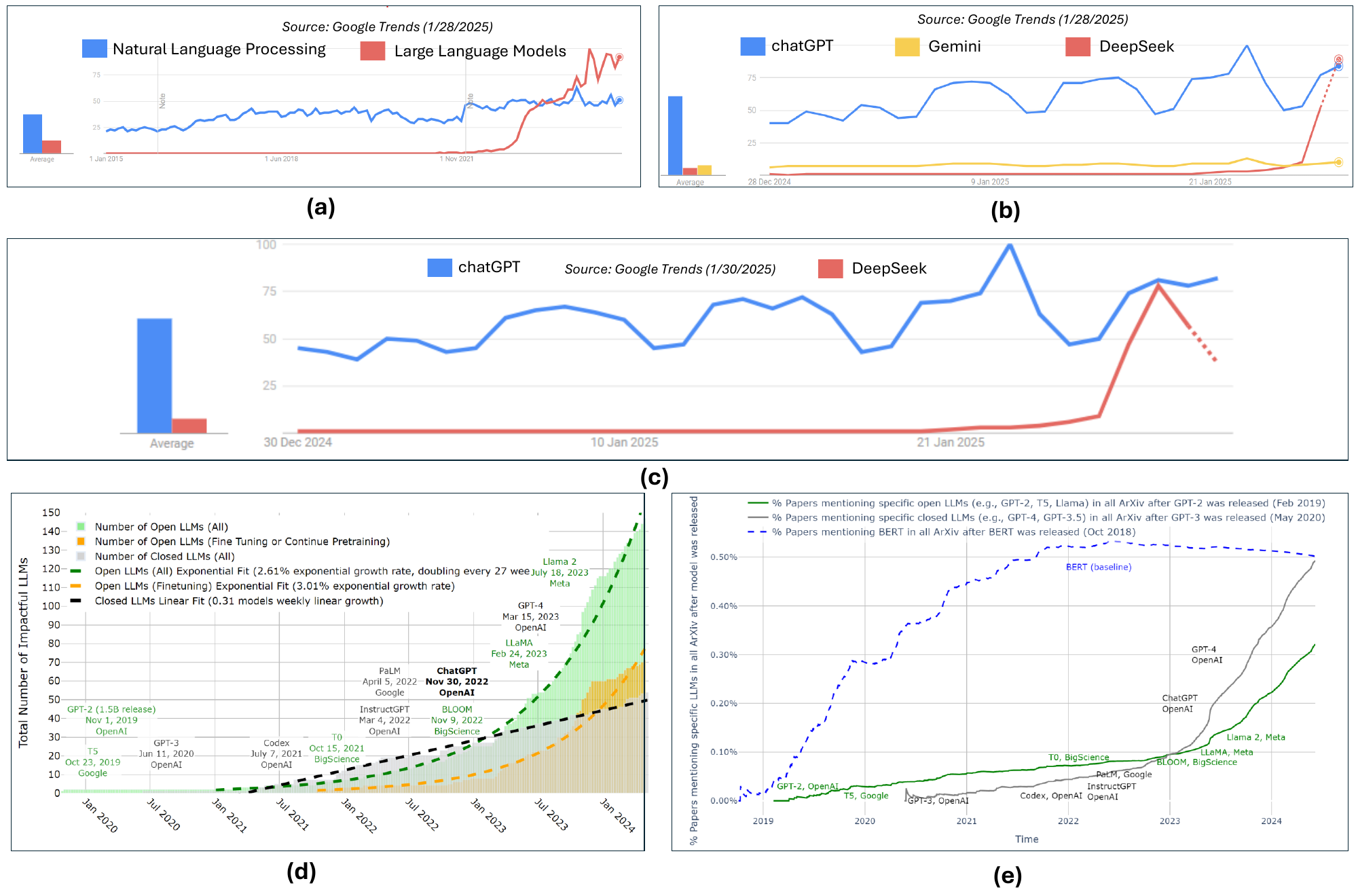}
\caption{\textbf{Analysis of NLP/LLM Interest}}
(a) Google Trends showing increasing interest in NLP and LLMs from 2015 to 2025; (b) Global interest for ChatGPT, Gemini, and DeepSeek on January 28 2025, highlighting DeepSeek's rapid rise; (c) ChatGPT and Deepseek global interest on January 30, 2025; (d) Growth rates of open sourced and close closed souced LLMs \cite{xu2025position} ; (e) Percentage of arXiv papers mentioning open LLMs or closed LLMs from 2019 onwards, with BERT as a baseline \cite{xu2025position}
\label{fig1}
\end{figure*}

Natural Language Processing (NLP) and Large Language Models (LLMs), including multimodal LLMs such as GPT-4o, DeepSeek-V2, and Gemini 1.5, have witnessed transformative advancements and significant growth in recent years, as illustrated by the surging global interest from both research and industry, as depicted in Figure \ref{fig1}a 

These technologies have become integral to systems and solutions across a diverse array of sectors, including healthcare \cite{cascella2023evaluating}, finance \cite{li2023large}, education \cite{neumann2024llm}, and entertainment\cite{qiu2024large}. Their remarkable capabilities in language understanding and generation have not only revolutionized these industries but have also spurred a new wave of innovation and application development \cite{weldon2024establishing, grant2025we}. Amidst this rapid expansion, the term “open-source” frequently surfaces within discussions about LLMs \cite{10.1145/3647782.3647803}. However, this descriptor is often misapplied or misunderstood. In many instances, developers may release only the model weights, that is, the trained parameters, without sharing the comprehensive suite of model assets such as model card, training data, code, sustainability factors (e.g., CO\textsubscript{2} emissions), or detailed development processes.  This gap is also widely discussed in the literature \cite{ramlochan2023openness} and in numerous tech blogs, including \cite{walker2024best}, to name a few.
% that are essential for a genuine open-source methodology  \cite{deitke2024molmo, white2024model}.

Although proprietary LLMs like OpenAI GPT-series (4/4o) \cite{achiam2023gpt} exhibit strong performance, their closed-source nature limits access to API-based interactions. In contrast, open-weight models like Meta LLama-series \cite{touvron2023llama} provide downloadable model weights under non-proprietary licenses, enabling specialized deployments and cost-effective fine-tuning. For instance, Princeton’s Llemma leverages Code Llama for advanced mathematical modeling \cite{azerbayev2023llemma}, showcases the flexibility and cost benefits of open-weight models.

The distinction between "open" and "closed" LLMs is evident in their adoption trends. Closed models like GPT-3 followed a linear growth pattern (gray bars, Figure \ref{fig1}d), while open LLMs surged after Meta’s Llama release, driving exponential adoption (green and orange bars, Figure \ref{fig1}d). Figure \ref{fig1}e further illustrates how open source models increasingly attract scientific focus compared to the same with proprietary models such as GPT-4. 

Despite this growing interest, the term “open-source” has frequently been used interchangeably with “open weights”, leading to confusion in discussions about model accessibility. Many models labeled as open-source provide access only to their trained weights while withholding essential components such as training data, fine-tuning methodologies, and full implementation details. This distinction is critical, as true open-source models enable not just inference but also full transparency and reproducibility in AI research. A recent case highlighting the confusion between open-source and open-weight models is DeepSeek-R1 \cite{guo2025deepseek}. Initially surpassing ChatGPT in search interest (Figure \ref{fig1}b), its popularity rapidly declined (Figure \ref{fig1}c), reflecting unmet expectations. While DeepSeek-R1 provides weights and partial code under the MIT license \footnote{\url{https://opensource.org/license/mit}}, it lacks full open-source transparency, including access to training data and methodologies. This partial openness, common to models like ChatGPT and Google’s Gemini, allows broader usage compared to fully closed models, but restricts deeper architectural modifications, evaluation of biases, and further enhancement of the training processes and datasets.

This ambiguity in AI terminologies necessitates clearer distinctions between open-source and open-weight models. True open-source AI requires full transparency, including training data and development processes, fostering reproducibility and ethical AI advancements. Defining and broadly adopting clear standards would enhance transparency, set realistic expectations, and promote responsible AI development.

\subsection{Aim and Objectives}
The goal of this study is to  critically examines transparency practices of such "open-weight" LLMs, using DeepSeek-R1 and ChatGPT4o as primary examples, to map the distinctions between open-weight and fully open-source models. By doing so, we aim to:
% This narrative review critically examines transparency practices of such "open-weight" LLMs, using DeepSeek-R1 and ChatGPT4o as primary examples, to map the distinctions between open-weight and fully open-source models. By doing so, we aim to:
\begin{itemize} 
\item Elucidate the terminological ambiguities surrounding "open-source" within the AI domain, specifically distinguishing between truly open-source models and those termed "open-weight" which offer limited transparency.
\item Investigate the implications of partial transparency on the reproducibility, community engagement, and ethical dimensions of AI development, emphasizing how these factors influence the practical deployment and trustworthiness of LLMs.
\item Propose clearer guidelines and standards to differentiate truly open-source methodologies and models from strategies that merely provide access to pre-trained model weights.
\end{itemize}

With this study, we seek to contribute to an informed advancement of responsible AI, where both technological innovation and collaborative transparency are harmonized. The following sections describe the current landscape of LLMs, the tensions between proprietary and open-weight models, and the broader impacts of these approaches on the AI research community.
\begin{figure*}[ht]
\centering
\includegraphics[width=0.60\linewidth]{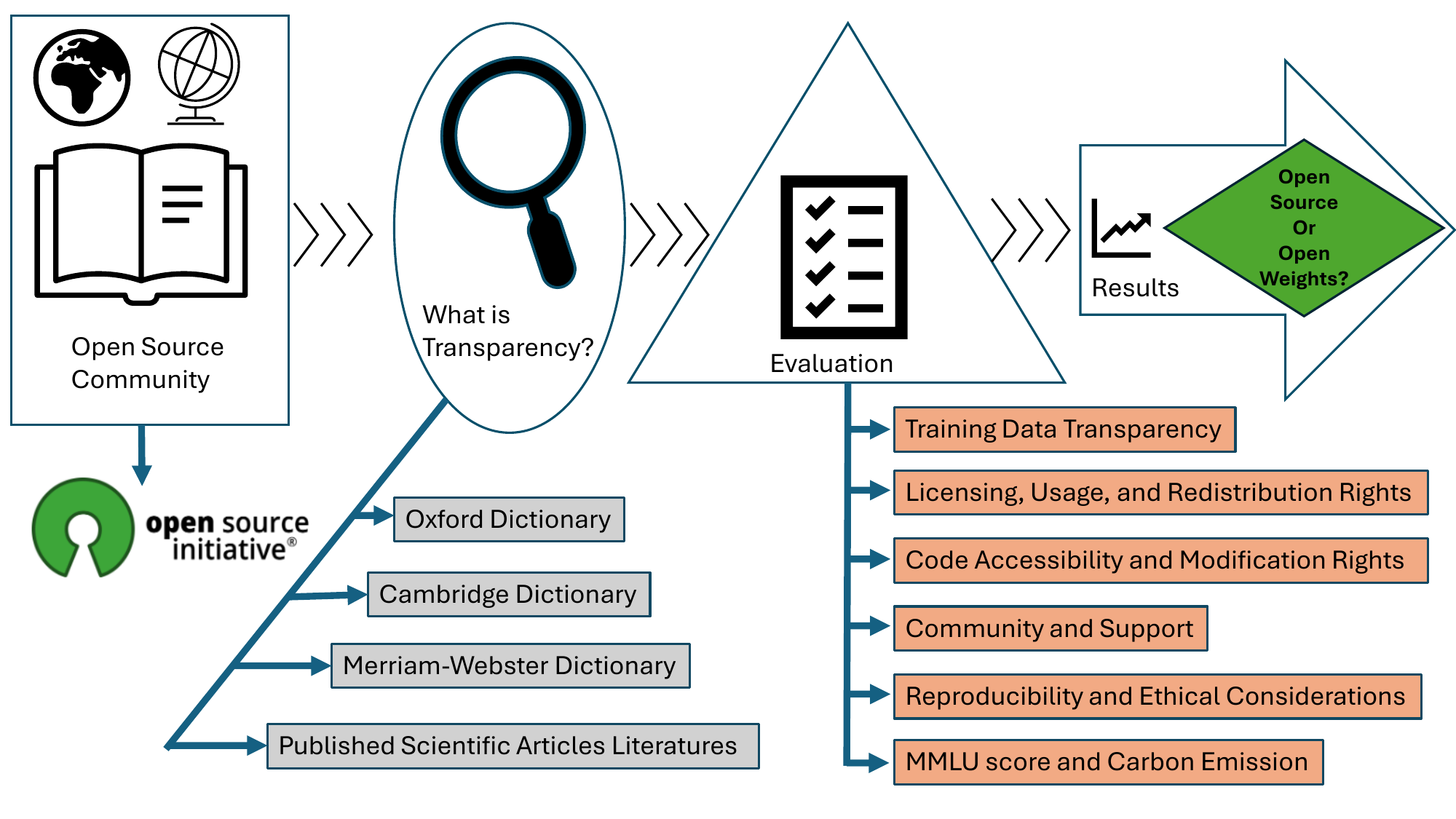}
\caption{Overview of the methodologies used in evaluating ChatGPT, DeepSeek, and SoTA multimodal LLMs}
\label{fig:methodology}
\end{figure*}
\section{Methodology}
 This study systematically examines the concepts of openness and transparency in the development and dissemination of LLMs. A multi-stage approach is used in this study,beginning with a thorough examination of foundational concepts and progressing through detailed analyses of licensing types and transparency definitions as they relate to AI systems.
%i remove narrative review , i think we overemphasizing, i think this study is more a scientific perspective than simple review

\subsection{Research Design}
\begin{figure}
\centering
\scalebox{0.60}{ 
\begin{tikzpicture}
  \path[mindmap,concept color=blue!50,text=black]
    node[concept] {Evaluating Transparency in Open Weight vs. Open Source Models}
    [clockwise from=0]
    child[concept color=green!50!black] {
      node[concept] {Factors Influencing Classification}
      [clockwise from=90]
      child { node[concept] {Efficiency in Deployment} }
      child { node[concept] {Scalability Challenges} }
      child { node[concept] {Operational Transparency} }
    }
    child[concept color=red!60] {
      node[concept] {Impact of Training Methodologies}
      [clockwise from=-30]
      child { node[concept] {Transparency in Training Processes} }
      child { node[concept] {Reproducibility Issues} }
    }
    child[concept color=orange!60] {
      node[concept] {Implications of Limited Data Sharing}
      [clockwise from=210]
      child { node[concept] {Understanding and Utilization} }
      child { node[concept] {Impact on Performance Metrics} }
      child { node[concept] {User and Developer Transparency} }
    };
\end{tikzpicture}
}
\caption{Illustration of an integrative mind map strategy developed to systematically evaluate transparency and accessibility of 112 LLMs from 2019 to the present. The diagram organizes critical dimensions, including factors influencing model classification, impacts of training methodologies, and consequences of limited data sharing—to comprehensively assess operational efficiency, scalability, and reproducibility challenges in open-weight versus open-source models.}
\label{fig:mindmap_transparency}
\end{figure}
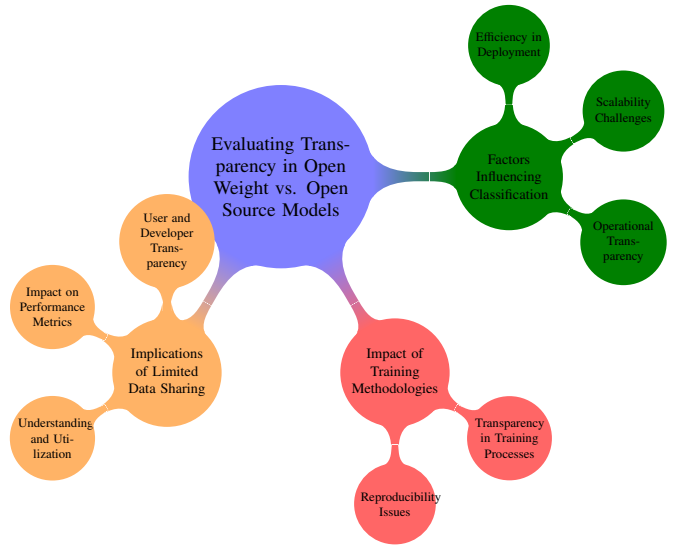

% As mentioned before, 
This study adopts a multi-stage research design to evaluate the openness and transparency of SoTA  LLMs. As illustrated in Figure \ref{fig:methodology}, the approach integrates established open-source criteria, foundational linguistic definitions of “transparency”, and an extensive review of scholarly AI literature. A concise mind map (Figure \ref{fig:mindmap_transparency}) further delineates the core analytical branches, structured into three research questions (RQs) guiding the study. Below, each methodological component is described in detail.

\begin{figure*}[ht]
\centering
\includegraphics[width=0.70\linewidth]{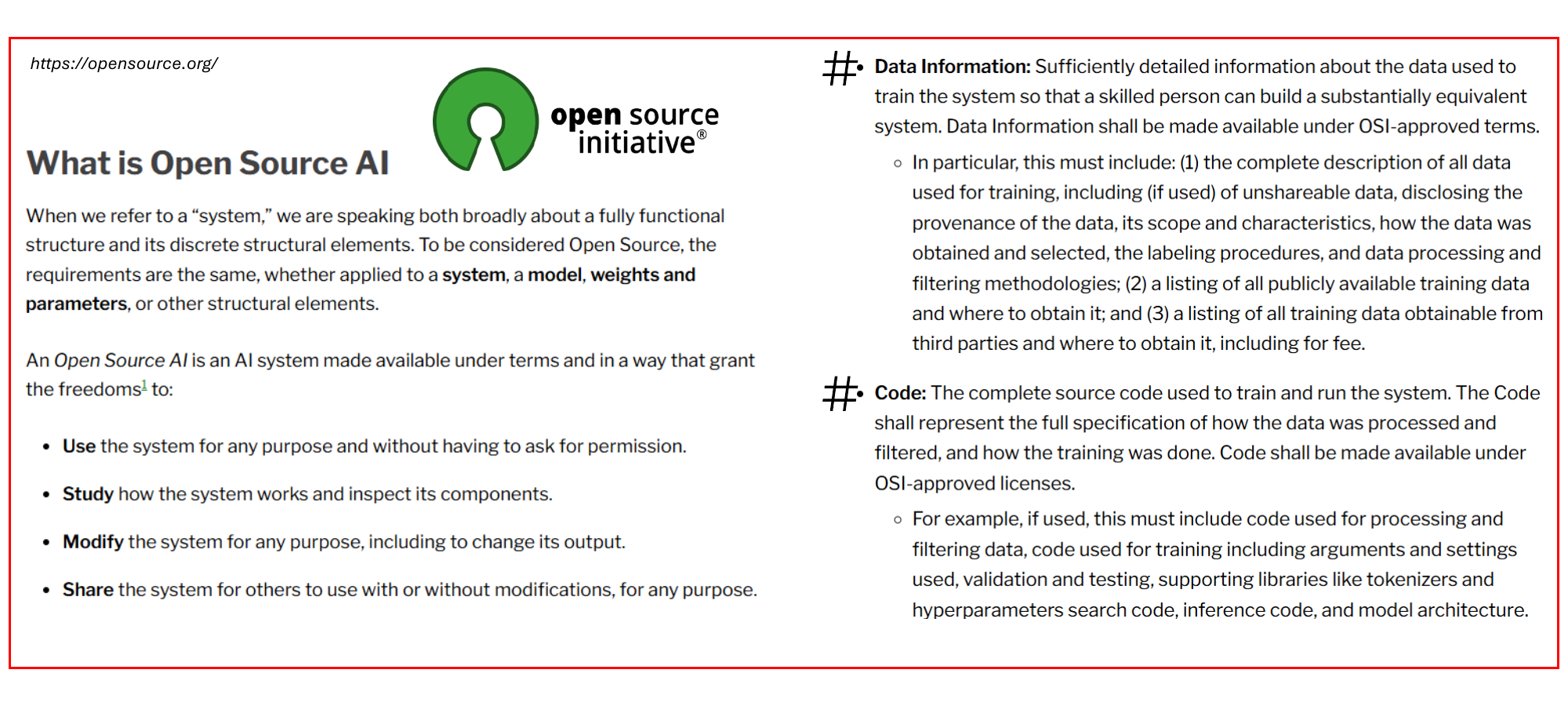}
\caption{OSI's first official release of the open source definition, which sets foundational criteria/attributes for Openness in AI}
\label{fig:osi}
\end{figure*}

\subsection{Criteria for Openness and Transparency}

\textbf{Open-Source LLMs:} An open-source LLM provides unrestricted access to its entire codebase, including the model architecture, training data, and the training processes \cite{ramlochan2023openness}. Beyond the code and weights, a truly open-source model also discloses key factors such as performance benchmarks, bias mitigation strategies, computational efficiency, and sustainability metrics  (e.g., Carbon dioxide emissions, energy consumption). For example, Meta’s LLamA aligns with the open-source paradigm by offering detailed insights into its design and implementation.

The primary goal with open-source models is to ensurecomplete transparency and flexibility. This openness enables comprehensive understanding, recreation, and reproducibility , even though some usage restrictions may still apply.  Such transparency allows the research community to scrutinize, improve, and tailor models for diverse applications. Developing and maintaining such models, however, demands substantial effort and resources, making the open-source approach both a technical and logistical challenge. For example, early models like  GPT-1 and GPT-2 were released as open-source projects, providing access to their training data, code and model weights. With subsequent versions like GPT-3, OpenAI shifted to a closed-source approach, restricting access to the model architecture, code, and weights. This trend continued with GPT-4, which also remains proprietary.

\textbf{Open-Weight LLMs:} Open-weight LLMs make their pre-trained model weights \cite{promptmetheusopenweights}, the parameters learned during the pre-training process, publicly available, while the underlying code, training data, or training methodologies may remain proprietary. Open-weight models, while more accessible and easier to deploy than closed-source models, do not provide the same level of insight into the model's inner workings as fully open-source models would. Meta’s LLama series is a prime example of an open-weight LLM. Researchers can download the pre-trained weights to fine-tune and deploy the model for various applications. However, while LLama weights are available, the full training pipeline, including the code and data, remains proprietary. This enables a balance between accessibility and intellectual property protection.

\subsubsection{Open Source and Licensing Types}
OSI stands for the Open Source Initiative \cite{opensource}. It is a non-profit organization dedicated to promoting and protecting open source software.  OSI is best known for its Open Source Definition (OSD), which outlines the criteria that a software license must meet to be considered "open source." These criteria include free redistribution, source code availability, the ability to create derivative works, and non-discrimination, among others. Essentially, OSI serves as a guardian of open source principles, ensuring that software labeled as open source truly adheres to standards that promote collaboration, transparency, and freedom in software development.

The primary attributes of the OSI's official definition of open-source AI are illustrated in Figure \ref{fig:osi}. OSI emphasizes that for an AI system to be truly open source, there must be unrestricted access to its entire structure. This means that key components—such as the model weights, source code, and training data—must be accessible under OSI-approved terms.  uch access allows any user to use, modify, share, and fully understand the AI system without needing special permissions.

Transparency refers to the clarity and understandability of the underlying mechanisms that drive AI systems. It is achieved when training data and code are available, enabling stakeholders to replicate and scrutinize the AI's decision-making processes \cite{larsson2020transparency, felzmann2020towards, von2021transparency}. This openness ensures that AI operations are not only visible but also comprehensible and accountable, thereby enhancing trust and fostering collaboration in AI development and application.

Open source software licenses further define the usage, modification, and distribution rights for software \cite{contractor2022behavioral}.
They are critical for both protecting creators and enabling users to innovate and adapt software to their needs \cite{quintais2023platforms}. For example, the MIT License, highly permissive, allows almost unrestricted use provided the original copyright is included.Similarly, the Apache License 2.0 \footnote{\url{https://www.apache.org/licenses/LICENSE-2.0}} permits broad use—including modifications and distributions—with the additional safeguard of patent rights protection
Although Creative Commons licenses \footnote{\url{https://creativecommons.org/share-your-work/cclicenses/}} are primarily designed for creative content, variants such as CC-BY-4.0 can also govern software use by allowing commercial use provided that proper credit is given to the creator.  Choosing the right license involves careful consideration of the intended use, attribution requirements, and legal protections, ensuring that software developers can support their objectives while fostering broader collaboration and innovation within the community. Table \ref{tab:license_comparison} provides an overview of popular licenses in AI practices, highlighting the varying degrees of permissiveness—from the flexible MIT License \footnote{\url{https://opensource.org/license/mit}} to the stricter copyleft provisions of the GNU GPL 3.0 \footnote{\url{https://www.gnu.org/licenses/gpl-3.0.en.html}}.

\begin{table}[H]
\centering
\caption{AI Licenses: A Comprehensive Comparison of Popular Types detailing their requirements for copyright preservation, 
patent grants, modification rights, distribution terms, and special clauses}
\label{tab:license_comparison}
\fontsize{8}{10}\selectfont
\begin{tabular}{%
  >{\raggedright\arraybackslash}m{2cm}%
  >{\centering\arraybackslash}m{1cm}%
  >{\raggedright\arraybackslash}m{1cm}%
  >{\raggedright\arraybackslash}m{1cm}%
  >{\raggedright\arraybackslash}m{1cm}%
  >{\centering\arraybackslash}m{1cm}%
}
\hline
\textbf{License Type} & \textbf{Copyright}\\[0.8ex]
 & \textbf{Preservation} & \textbf{Patent Grant} & \textbf{Modification Rights} & \textbf{Distribution Terms} & \textbf{Special Clauses} \\[1ex]
\hline 
MIT License \cite{mitlicense} & Required & No explicit grant & Unlimited modifications & Must include original notices & - \\[1ex]
Apache License 2.0 \cite{apachelicense} & Required & Includes patent rights & Modifications documented & Must include original notices & - \\[1ex]
GNU GPL 3.0 \cite{gplv3} & Required & - & Derivative works must also be open source & Source code must be disclosed & Strong copyleft \\[1ex]
BSD License \cite{bsdlicense} & Required & No explicit grant & Unlimited modifications & No requirement to disclose source & No endorsement \\[1ex]
Creative ML OpenRAIL-M \cite{openrail-m} & Required & - & Ethical use guidelines & Must include original notices & Ethical guidelines \\[1ex]
CC-BY-4.0 \cite{ccby4} & Credit required & - & Commercial and non-commercial use allowed & Must credit creator & - \\[1ex]
CC-BY-NC-4.0 \cite{ccbync4} & Credit required & - & Only non-commercial use allowed & Must credit creator & Non-commercial use only \\[1ex]
BigScience OpenRAIL-M \cite{bigscienceopenrail} & Required & - & Ethical use guidelines & Must include original notices & Ethical guidelines \\[1ex]
BigCode Open RAIL-M v1 \cite{bigcodeopenrail} & Required & - & Ethical use guidelines & Must include original notices & Ethical guidelines \\[1ex]
Academic Free License v3.0 \cite{afl3} & Required & Includes patent rights & Unlimited modifications & Must include original notices & - \\[1ex]
Boost Software License 1.0 \cite{boost1} & Required & No explicit grant & Unlimited modifications & Must include original notices & - \\[1ex]
BSD 2-clause “Simplified” \cite{bsd2} & Required & No explicit grant & Unlimited modifications & No requirement to disclose source & No endorsement \\[1ex]
BSD 3-clause “New” or “Revised” \cite{bsd3} & Required & No explicit grant & Unlimited modifications & No requirement to disclose source & No endorsement \\[1ex]
\hline
\end{tabular}
\end{table}

\subsubsection{Open Source and Transparency}
Following the OSI guidelines, dictionary definitions further support the concept of open source and transparency. \textit{According to Oxford  \footnote{\url{https://www.oed.com/}}  , open source software is described as ``Used to describe software for which the original source code is made available to anyone.'' Cambridge further explains that open source software or information can be ``obtained legally and for free from the internet, and can be used, shared or changed without paying or asking for special permission.'' Merriam-Webster defines it as ``Having the source code freely available for possible modification and redistribution.'' For transparency, Oxford states it as ``The quality of something, such as glass, that allows you to see through it.'' Cambridge calls it ``The characteristic of being easy to see through.'' Merriam-Webster describes transparency as ``The quality or state of being transparent so that bodies lying beyond are seen clearly.''}
These definitions set a foundational understanding to evaluate the transparency practices in AI systems, as shown in Table \ref{tab:transparency-definitions}, which presents a literature review and definitions derived from 10 popular literature defining transparency in AI systems.

\begin{table}[ht]
\raggedleft  % Makes the table flush left
\caption{Unified Definitions of Transparency in AI from the published literature.}
\scriptsize
\begin{tabular}{p{2cm} p{4.5cm}}
\textbf{Author and Reference} & \textbf{Definition} \\[1ex]
\hline
Lipton, Z. C. (2018). \cite{lipton2018mythos}  & “Transparency in machine learning models means understanding how predictions are made, underscored by the \textbf{availability of training datasets and code}, which supports both local and global interpretability.” \\[1ex]
Doshi-Velez, F., \& Kim, B. (2017). \cite{doshi2017towards} & “Transparency in AI refers to the ability to understand and trace the decision-making process, including the \textbf{availability of training datasets and code}. This enhances the clarity of how decisions are made within the model.” \\[1ex]
Arrieta, A. B., et al. (2020). \cite{arrieta2020explainable} & “AI transparency means understanding the cause of a decision, supported by the \textbf{availability of training datasets and code}, which fosters trust in the AI's decision-making process.” \\[1ex]
Ribeiro, M. T., Singh, S., \& Guestrin, C. (2016). \cite{ribeiro2016should} & “Transparency in AI models provides insights into model behavior, heavily reliant on the \textbf{availability of training datasets and code} to illuminate how input features influence outputs.” \\[1ex]
Goodman, B., \& Flaxman, S. (2017). \cite{goodman2017european} & “Transparency involves scrutinizing the algorithms and data used in decisions, emphasizing the \textbf{availability of training datasets and code} to ensure fairness and accountability.” \\[1ex]
Molnar, C. (2020). \cite{molnar2020interpretable} & “Transparency in AI refers to clear communication about decision-making processes, facilitated by the \textbf{availability of training datasets and code}, allowing for better understanding of model outputs.” \\[1ex]
Rudin, et al. (2021). \cite{rudin2022interpretable} & “Transparency is offering clear, interpretable explanations for decisions, which necessitates the \textbf{availability of training datasets and code} for full interpretability.” \\[1ex]
Bhatt, et al. (2020). \cite{bhatt2020explainable} & “Transparency involves making AI's decision-making process accessible, underlined by the \textbf{availability of training datasets and code}, aligning with ethical standards.” \\[1ex]
Gilpin, et al. (2021). \cite{gilpin2018explaining} & “Transparency ensures clear explanations of model behavior, significantly relying on the \textbf{availability of training datasets and code} for technical and operational clarity.” \\
\hline
\end{tabular}
\label{tab:transparency-definitions}
\end{table}

\subsection{ Synthesis of Literature}

\subsubsection{Search Strategy}
The study first identified the requirements outlined by the OSI \footnote{\url{https://opensource.org/}} as the baseline for evaluating AI models. These criteria covers various facets of openness, including licensing provisions, access to source code, free redistribution rights, and the ability to modify or derive new work/models from the original codebase. 
Building on the OSI standards, the concept of “transparency” was clarified through an examination of widely used dictionaries (Oxford, Cambridge, and Merriam-Webster) \cite{rottger2024safetyprompts}.
% A narrative synthesis approach was adopted to integrate the broad spectrum of scholarship on AI openness and transparency. 
Key steps included:

\textbf{Databases and Sources:} The selection of databases was aligned with the goal of capturing a breadth of interdisciplinary research that intersects with artificial intelligence. Academic repositories such as ACM Digital Library, IEEE Xplore, Elsevier, Nature, Scopus, ScienceDirect, SpringerLink, Wiley Online Library, MathSciNet and renowned pre-print servers like arXiv were chosen for their extensive coverage of both technical and ethical dimensions pertinent to AI. These platforms are renowned for their consolidation of high-impact and specialized journals, which provide critical insights into both emerging and established research areas within technology and applied sciences.

Our literature search was further reinforced by prioritizing papers that are highly cited within the academic community. Citation counts, often seen as a proxy for the influence and relevance of a study, were utilized as a key metric in selecting sources. Papers with exceptionally high citation counts (e.g., > 3000 citations), were specifically targeted. This criterion was instrumental because highly cited papers typically reflect pivotal developments in the field and are often the genesis of new research trajectories or shifts in scientific paradigms. The search terms used were ``Transparency in AI'', ``Transparency in LLMs'', ``Explainable AI'', ``Reproducible AI'', ``Open Source AI'', ``Open Source Model'', ``Open Source Software'', ``Fairness in AI'', ``Ethical AI'', ``Responsible AI'', ``Bias in AI'', ``Sustainable AI'', ``Green AI'', ``AI Ethics'', ``AI Accountability'', ``Interpretable AI'', ``AI Robustness'', ``AI Reliability'', and ``AI Compliance''.

\textbf{Timeframe} The literature selected for this study spans publications from 2017 onward—a timeframe strategically chosen to align with the introduction of Transformers. In 2017, Vaswani et al. published Attention is All You Need \cite{vaswani2017attention}, marking the beginning of a new era in AI by introducing a model architecture based on attention mechanisms. Following this, the launch of GPT-2, T5, BART, and several other language model architectures further advanced the field, shaping the development of modern LLMs. We systematically assessed these models to identify models that exemplify various degrees of openness, including open-source and open-weight practices.

\begin{figure}[ht]
\centering
\includegraphics[width=0.97\linewidth]{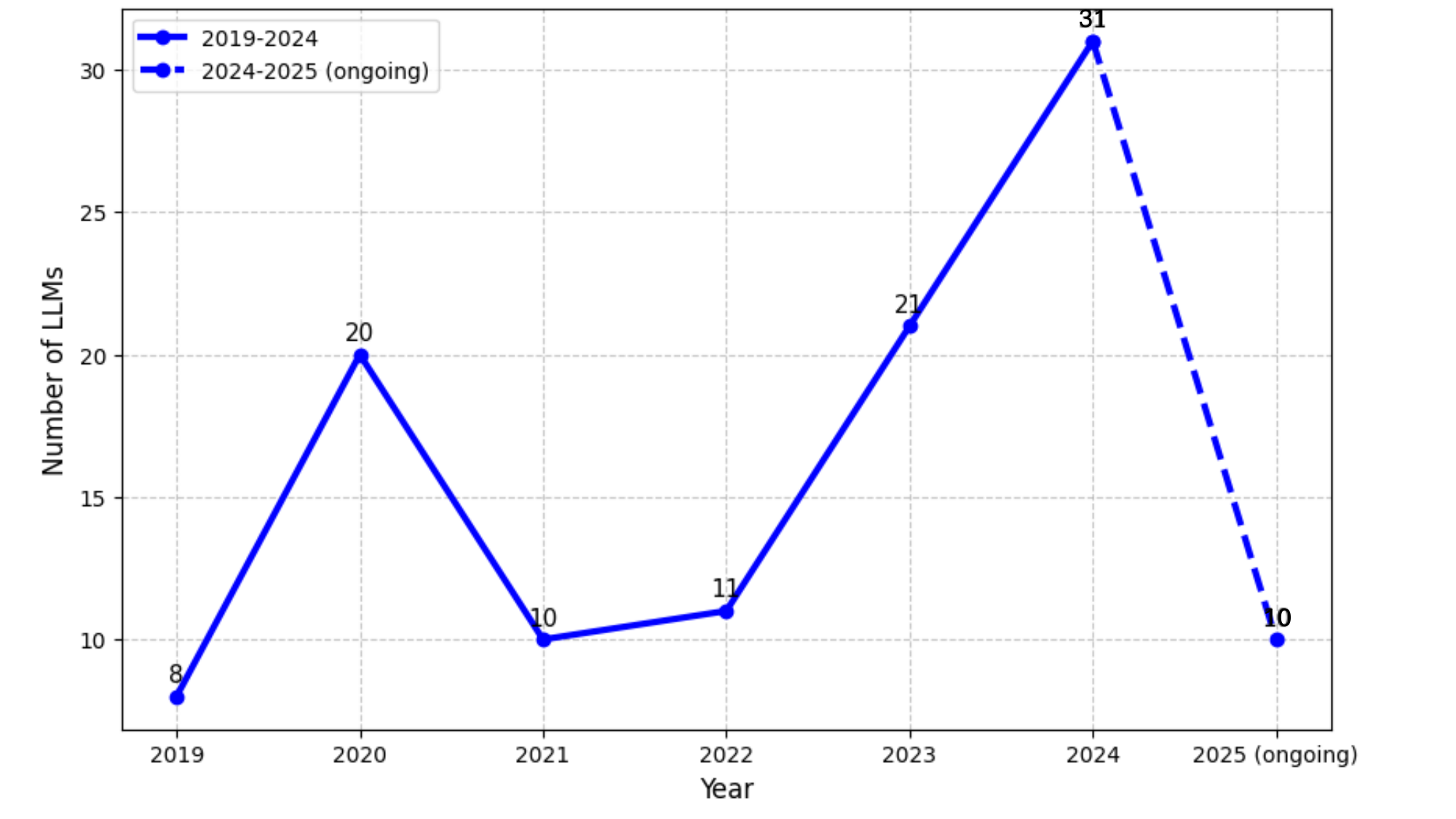}
\caption{Temporal distribution of 112 research papers analyzed in this study, spanning from 2019 to 2025. The plot reveals a steadily increasing trend in LLM studies, underscoring rapid advancements in transparency and accessibility.}
\label{fig:papersTotal}
\end{figure}

In the process of synthesizing these findings, we evaluated a total of 112 LLMs, a sample that represents the diverse and rapidly evolving landscape of language models from 2019 to 2025. These models were analyzed based on a wide array of architectural specifications—such as the number of layers, hidden unit sizes, attention head counts, and overall parameter scales—as well as openness metrics including licensing type and the public availability of training resources. The model development trend, illustrated in Figure~\ref{fig:papersTotal}, provides a visual representation of the evolution of these models. The figure shows that although the foundational literature for LLMs was established with the advent of Transformers in 2017, the major model breakthroughs and integrated transparency and accessibility features have predominantly materialized from 2019 onward and more post-ChatGPT era (Nov. 2022).

\textbf{Inclusion Criteria} A thorough literature review was conducted to locate transparency within broader discourses in AI development and ethics. This review captured highly cited articles and technical reports, emphasizing themes such as explainable AI, reproducibility, interpretability, and responsible AI governance \cite{raza2025responsible}. By synthesizing these studies, our study addressed both technical (e.g., code-level transparency) and ethical (e.g., data biases) dimensions of openness.

\subsection{Evaluation Framework and Application}
Findings from the previous stages were synthesized into five key dimensions representing critical facets of open-source and open-weight classifications:
\begin{enumerate}
    \item Licensing, Usage, and Redistribution Rights
    \item Code Accessibility and Modification Rights
    \item Training Data Transparency
    \item Community and Support
    \item MMLU Score and Carbon Emissions
    \item Ethical Considerations and Reproducibility
\end{enumerate}

Each of these dimensions was assessed to determine whether a given model adhered to OSI-like openness or employed more restrictive practices similar to “open-weight” approaches (i.e., sharing only the model parameters). SoTA LLMs were systematically evaluated against these five dimensions as 1) Licensing, usage, and redistribution rights , 2) Training Code and Training Data, 3) Community Support, 4) Open source, and 5) Open Weights. Any evidence of collaborative contributions or transparent reporting of potential biases and vulnerabilities was also documented.

\subsection{Research Questions}
The methodology section of this study was structured around a detailed mind map, as depicted in Figure \ref{fig:mindmap_transparency}. This visual representation, employed to assess the transparency and openness of SoTA multimodal LLMs, organized the analytical framework into three main branches, each corresponding to a specific research question (RQ) as follows:

\begin{enumerate}
    \item What drives the classification of LLMs as \textit{open weights} rather than \textit{open source}, and what impact do these factors have on their efficiency and scalability in practical applications?
    \item How do current training approaches influence transparency and reproducibility, potentially prompting developers to favor open-weight models?
\item How does the limited disclosure of training data and methodologies impact both the performance and practical usability of these models, and what future implications arise for developers and end-users?
 \end{enumerate}

This methodology integrates well-established open-source standards, linguistically and ethically grounded definitions of transparency, and a structured evaluation framework. The outcome is an assessment of whether leading MLLMs adhere to open-source principles or merely present limited transparency through open-weight practices. The subsequent sections detail the findings that emerged from applying this framework, highlighting significant discrepancies and implications for researchers, developers, and broader AI stakeholders.

\section{Results}
\label{results}
\subsection{Overall Findings on Openness and Transparency}
Drawing on OSI guidelines, dictionary-based definitions of transparency, and scholarly literature, this narrative review reveals that many models marketed or perceived as “open” primarily provided open weights (i.e., publicly available trained parameters) rather than full open-source access (i.e., source code, training data, and detailed methodologies). Table~\ref{tab:summary_models} outlines these distinctions across leading multimodal LLMs.

This comprehensive table (Table ~\ref{tab:summary_models}) compares 112 LLMs released between 2019 and 2025 in terms of release year, training data, and other key features. Early models, such as GPT-2 \citep{brown2020language} and BERT \citep{devlin2019bertpretrainingdeepbidirectional}, primarily focused on foundational capabilities including improved text generation, masked language modeling, and next-sentence prediction. These models relied on relatively simple training data and featured basic natural language processing tasks. However, subsequent developments have led to a remarkable progression in both complexity and functionality. Recent models—such as DeepSeek-R1 \citep{guo2025deepseek} and advanced iterations of ChatGPT, introduce enhanced multimodal capabilities, advanced reasoning through mixture-of-experts (MoE) architectures, and efficient scaling strategies. The table demonstrates that models released after 2020 increasingly leverage diverse and massive training datasets, from extensive web corpora to hybrid synthetic-organic data—which significantly boost performance. Moreover, these models exhibit notable improvements in precision, processing speed, and bias mitigation. Although many SoTA models disclose only pre-trained weights, thereby limiting reproducibility, an emerging trend toward greater transparency regarding training methodologies was observed. This evolution reflects an industry-wide shift towards balancing commercial interests with greater accountability and openness in AI research.

% Column setup
\newcolumntype{M}{>{\raggedright\arraybackslash}p{2.5cm}} % Model+Year+Citation
\newcolumntype{D}{>{\raggedright\arraybackslash}p{2cm}}   % Training Data
\newcolumntype{K}{>{\raggedright\arraybackslash}p{2cm}}   % Key Features

\tablefirsthead{%
\toprule
\multicolumn{1}{M}{\textbf{Model, Year \& Citation}} & 
\multicolumn{1}{D}{\textbf{Training Data}} & 
\multicolumn{1}{K}{\textbf{Key Features}} \\
\midrule
}
\tablehead{%
\multicolumn{3}{c}{\small\sl Continued from previous column/page}\\
\toprule
\multicolumn{1}{M}{\textbf{Model, Year \& Citation}} & 
\multicolumn{1}{D}{\textbf{Training Data}} & 
\multicolumn{1}{K}{\textbf{Key Features}} \\
\midrule
}
\tabletail{\midrule\multicolumn{3}{c}{\small\sl Continued on next column/page}\\}
\tablelasttail{\bottomrule}

{\scriptsize
\tablecaption{\textbf{Detailed specifications of Large Language Models (2019--2025), including the model name and citation, release year, training data characteristics, and key features. It offers a comparative analysis across these crucial aspects, providing insights into the evolution, dataset diversity, and unique capabilities of each model}. \label{tab:summary_models}}%
\begin{supertabular}{M D K}
1.GPT-2 (2019) \cite{brown2020language} & WebText dataset (8M web pages) & Improved text generation, zero-shot learning \\
2.Legacy ChatGPT-3.5 (2022) & Text/Code (pre-2021) & Basic text tasks, translation \\
3.Default ChatGPT-3.5 (2023) & Text/Code (pre-2021) & Faster, less precise \\
4.GPT-3.5 Turbo (2023) & Text/Code (pre-2021) & Optimized accuracy \\
5.ChatGPT-4 (2023) & Text/Code (pre-2023) & Multimodal (text), high precision \\
6.GPT-4o (2024) \cite{hurst2024gpt} & Text/Code (pre-2024) & Multimodal (text/image/audio/video) \\
7.GPT-4o mini (2024) & Text/Code (pre-2024) & Cost-efficient, 60\% cheaper \\
8. o1-preview \cite{jaech2024openai} (2024) & STEM-focused data & System 2 thinking, PhD-level STEM \\
9. o1-mini (2024) & STEM-focused data & Fast reasoning, 65K tokens output \\
10. o1 (2025) & General + STEM data & Full o1 reasoning, multimodal \\
11. o1 pro mode (2025) & General + STEM data & Enhanced compute, Pro-only \\
12. o3-mini (2025) & General + STEM data & o1-mini successor \\
13. o3-mini-high (2025) & General + STEM data & High reasoning effort \\
14. DeepSeek-R1 \cite{guo2025deepseek} (2025) & Hybrid dataset of 9.8T tokens from synthetic and organic sources & Mixture of Experts (MoE), enhanced with mathematical reasoning capabilities \\
15. DeepSeek LLM \cite{bi2024deepseek} (2023) & Books+Wiki data up to 2023 & Scaling Language Models \\
16. DeepSeek LLM V2 \cite{liu2024deepseek} (2023) & Highly efficient training & MLA, MoE, Lowered costs \\
17. DeepSeek Coder V2 \cite{zhu2024deepseek} (2023) & Supports 338 languages & Enhanced coding capabilities \\
18. DeepSeek V3 \cite{liu2024deepseek3} (2023) & Advanced MoE architecture & High-performance, FP8 training \\
19. BERT-Base \cite{devlin2019bertpretrainingdeepbidirectional} (2019) & Books+Wiki data collected up to 2019 & Masked Language Modeling (MLM) \\
20. BERT-Large \cite{devlin2019bertpretrainingdeepbidirectional} (2019) & Books+Wiki data collected up to 2019 & Next Sentence Prediction (NSP) \\
21. T5-Small \cite{raffel2020exploring} (2020) & C4 (Large-scale text dataset) & Text-to-text, encoder-decoder \\
22. T5-Base \cite{raffel2020exploring} (2020) & C4 (Large-scale text dataset) & Text-to-text, scalable, encoder-decoder \\
23. T5-Large \cite{raffel2020exploring} (2020) & C4 (Large-scale text dataset) & Text-to-text, scalable, encoder-decoder \\
24. T5-3B \cite{raffel2020exploring} (2020) & C4 (Large-scale text dataset) & Text-to-text, scalable, encoder-decoder \\
25. T5-11B \cite{raffel2020exploring} (2020) & C4 (Large-scale text dataset) & Text-to-text, scalable, encoder-decoder \\
26. Mistral 7B \cite{jiang2023mistral} (2023) & Compiled from diverse sources totaling 2.4T tokens & Sliding Window Attention (SWA) \\
27. LLaMA 2 70B \cite{touvron2023llama} (2023) & Diverse corpus aggregated up to 2T tokens & Grouped Query Attention (GQA) \\
28. CriticGPT \cite{mcaleese2024llm} (2024) & Human feedback data & Fine-tuned for critique generation \\
29. Olympus (2024) & 40T tokens & Large-scale, proprietary model \\
30. HLAT \cite{fan2024hlat} (2024) & Not specified & High-performance, task-specific \\
31. Multimodal-CoT \cite{zhang2023multimodal} (2023) & Multimodal datasets & Chain-of-Thought reasoning for multimodal tasks \\
32. AlexaTM 20B \cite{soltan2022alexatm} (2022) & Not specified & Multilingual, task-specific \\
33. Chameleon \cite{team2024chameleon} (2024) & 9.2T tokens & Multimodal, high-performance \\
34. Llama 3 70B \cite{metaIntroducingMeta} (2024) & 2T tokens & High-performance, open-source \\
35. LIMA \cite{zhou2024lima} (2024) & Not specified & High-performance, task-specific \\
36. BlenderBot 3x \cite{xu2023improving} (2023) & 300B tokens & Conversational AI, improved reasoning \\
37. Atlas \cite{izacard2023atlas} (2023) & 40B tokens & High-performance, task-specific \\
38. InCoder \cite{fried2022incoder} (2022) & Not specified & Code generation, task-specific \\
39. 4M-21 \cite{bachmann20244m} (2024) & Not specified & High-performance, task-specific \\
40. Apple On-Device model \cite{mehta2024openelm} (2024) & 1.5T tokens & On-device, task-specific \\
41. MM1 \cite{mckinzie2024mm1} (2024) & 2.08T tokens & Multimodal, high-performance \\
42. ReALM-3B \cite{moniz2024realm} (2024) & 134B tokens & High-performance, task-specific \\
43. Ferret-UI \cite{you2024ferret} (2024) & 2T tokens & Multimodal, high-performance \\
44. MGIE \cite{fu2023guiding} (2023) & 2T tokens & Multimodal, high-performance \\
45. Ferret \cite{you2023ferret} (2023) & 2T tokens & Multimodal, high-performance \\
46. Nemotron-4 340B \cite{adler2024nemotron} (2024) & 9T tokens & High-performance, task-specific \\
47. VIMA \cite{jiang2023vima} (2023) & Not specified & Multimodal, high-performance \\
48. Retro 48B \cite{wang2023instructretro} (2023) & 1.2T tokens & High-performance, task-specific \\
49. Raven \cite{huang2023raven} (2023) & 40B tokens & High-performance, task-specific \\
50. Gemini 1.5 \cite{reid2024gemini} (2024) & Not specified & Multimodal, high-performance \\
51. Med-Gemini-L 1.0 \cite{saab2024capabilities} (2024) & 30T tokens & Medical-focused, high-performance \\
52. Hawk \cite{de2024griffin} (2024) & 300B tokens & High-performance, task-specific \\
53. Griffin \cite{de2024griffin} (2024) & 300B tokens & High-performance, task-specific \\
54. Gemma \cite{team2024gemma} (2024) & 6T tokens & High-performance, task-specific \\
55. Gemini 1.5 Pro \cite{reid2024gemini} (2024) & 30T tokens & Multimodal, high-performance \\
56. PaLi-3 \cite{chen2023pali} (2023) & Not specified & Multimodal, high-performance \\
57. RT-X \cite{padalkar2023open} (2023) & Not specified & Robotics-focused, high-performance \\
58. Med-PaLM M \cite{tu2024towards} (2024) & 780B tokens & Medical-focused, high-performance \\
59. MAI-1 \cite{MAI2024} (2024) & 10T tokens & High-performance, task-specific \\
60. YOCO \cite{sun2024you} (2024) & 1.6T tokens & High-performance, task-specific \\
61. phi-3-medium \cite{abdin2024phi} (2024) & 4.8T tokens & High-performance, task-specific \\
62. phi-3-mini \cite{abdin2024phi} (2024) & 3.3T tokens & High-performance, task-specific \\
63. WizardLM-2-8x22B \cite{FACE_2015} (2024) & Not specified & High-performance, task-specific \\
64. WaveCoder-Pro-6.7B \cite{yu2023wavecoder} (2023) & 20B tokens & Code-focused, high-performance \\
65. WaveCoder-Ultra-6.7B \cite{yu2023wavecoder} (2023) & 20B tokens & Code-focused, high-performance \\
66. WaveCoder-SC-15B \cite{yu2023wavecoder} (2023) & 20B tokens & Code-focused, high-performance \\
67. OCRA 2 \cite{mitra2023orca} (2023) & Not specified & High-performance, task-specific \\
68. Florence-2 \cite{xiao2024florence} (2024) & 5.4B visual annotations & Multimodal, high-performance \\
69. Qwen \cite{bai2023qwen} (2023) & 3T tokens & High-performance, task-specific \\
70. SeaLLM-13b \cite{nguyen2023seallms} (2023) & 2T tokens & Multilingual, high-performance \\
71. Grok-1 \cite{xai2024grok1} (2024) & 13.2T tokens & Incorporates humor-enhancing algorithms \\
72. Phi-4 \cite{abdin2024phi} (2024) & 9.8T tokens & Optimized for STEM applications \\
73. Megatron-LM \cite{shoeybi2019megatron} (2020) & Common Crawl, Wikipedia, Books & Large-scale parallel training, optimized for NVIDIA GPUs \\
74. Turing-NLG \cite{smith2022using} (2020) & Diverse web text & High-quality text generation, used in Microsoft products \\
75. CTRL \cite{keskar2019ctrl} (2019) & Diverse web text with control codes & Controlled text generation using control codes \\
76. XLNet \cite{yang2019xlnet} (2019) & BooksCorpus, Wikipedia, Giga5, ClueWeb & Permutation-based training, outperforms BERT on many benchmarks \\
77. RoBERTa \cite{liu2019roberta} (2019) & BooksCorpus, Wikipedia, CC-News, OpenWebText & Improved BERT with better pretraining techniques \\
78. ELECTRA \cite{clark2020electra} (2020) & BooksCorpus, Wikipedia & Replaces masked language modeling with a more efficient discriminative task \\
79. ALBERT \cite{lan2019albert} (2019) & BooksCorpus, Wikipedia & Parameter reduction techniques for efficient training \\
80. DistilBERT \cite{sanh2019distilbert} (2019) & BooksCorpus, Wikipedia & Distilled version of BERT, smaller and faster \\
81. BigBird \cite{zaheer2020big} (2020) & BooksCorpus, Wikipedia, PG-19 & Sparse attention mechanism for handling long sequences \\
82. Gopher \cite{rae2021scaling} (2021) & MassiveText dataset (2.5T tokens) & Focused on scaling laws and model performance \\
83. Chinchilla \cite{hoffmann2022training} (2022) & MassiveText dataset (1.4T tokens) & Optimized for compute-efficient training \\
84. PaLM \cite{chowdhery2023palm} (2022) & Diverse web text, books, code & Pathways system for efficient training, multilingual support \\
85. OPT \cite{zhang2022opt} (2022) & Diverse web text & Open-source alternative to GPT-3 \\
86. BLOOM \cite{workshop2022bloom} (2022) & ROOTS corpus (1.6T tokens) & Multilingual, open-source, collaborative effort \\
87. Jurassic-1 \cite{lieber2021jurassic} (2021) & Diverse web text & High-quality text generation, API-based access \\
88. Codex \cite{chen2021evaluating} (2021) & Code repositories (e.g., GitHub) & Specialized in code generation and understanding \\
89. T0 \cite{sanh2021multitask} (2021) & Diverse NLP datasets & Zero-shot task generalization \\
90. UL2 \cite{tay2022ul2} (2022) & Diverse web text & Unified pretraining for diverse NLP tasks \\
91. GLaM \cite{du2022glam} (2021) & Diverse web text & Sparse mixture of experts (MoE) architecture \\
92. ERNIE 3.0 \cite{sun2021ernie} (2021) & Chinese and English text & Knowledge-enhanced pretraining \\
93. GPT-NeoX \cite{black2022gpt} (2022) & The Pile (825GB dataset) & Open-source, large-scale, efficient training \\
94. CodeGen \cite{nijkamp2022codegen} (2022) & Code repositories (e.g., GitHub) & Specialized in code generation \\
95. FLAN-T5 \cite{chung2024scaling} (2022) & Diverse NLP datasets & Instruction fine-tuning for better generalization \\
96. mT5 \cite{xue2020mt5} (2020) & mC4 dataset (101 languages) & Multilingual text-to-text transfer \\
97. Reformer \cite{kitaev2020reformer} (2020) & Diverse web text & Efficient attention mechanism for long sequences \\
98. Longformer \cite{beltagy2020longformer} (2020) & BooksCorpus, Wikipedia & Efficient attention for long documents \\
99. DeBERTa \cite{he2020deberta} (2021) & BooksCorpus, Wikipedia & Disentangled attention mechanism \\
100. T-NLG \cite{TURING2020} (2020) & Diverse web text & High-quality text generation \\
101. Switch Transformer \cite{fedus2022switch} (2021) & Diverse web text & Sparse mixture of experts (MoE) \\
102. WuDao 2.0 \cite{waoDao2021} (2021) & Chinese and English text & Largest Chinese language model \\
103. LaMDA \cite{thoppilan2022lamda} (2021) & Diverse dialogue data & Specialized in conversational AI \\
104. MT-NLG \cite{smith2022using} (2021) & Diverse web text & High-quality text generation \\
105. GShard \cite{lepikhin2020gshard} (2020) & Diverse web text & Sparse mixture of experts (MoE) \\
106. T5-XXL \cite{raffel2020exploring} (2020) & C4 dataset & Large-scale text-to-text transfer \\
107. ProphetNet \cite{qi2020prophetnet} (2020) & BooksCorpus, Wikipedia & Future token prediction for better sequence modeling \\
108. DialoGPT \cite{zhang2019dialogpt} (2020) & Reddit dialogue data & Specialized in conversational AI \\
109. BART \cite{lewis2019bart} (2020) & BooksCorpus, Wikipedia & Denoising autoencoder for text generation \\
110. PEGASUS \cite{zhang2020pegasus} (2020) & C4 dataset & Pre-training with gap-sentences for summarization \\
111. UniLM \cite{dong2019unified} (2020) & BooksCorpus, Wikipedia & Unified pre-training for NLU and NLG tasks \\
112. Grok 3 (2025) & Synthetic Data & trained with ten times more computing power than its predecessor, Grok 2\\
\end{supertabular}
}

In recent years, LLMs have been further advanced through continuous refinement of critical features essential for practical applications. Models such as T5-XXL \citep{raffel2020exploring} have significantly expanded both the scale and diversity of training data, transitioning from datasets with millions of tokens to those with trillions. This dramatic increase in training volume has enabled improvements in computational efficiency, reproducibility, and bias mitigation. Additionally, evolving training methodologies—from basic text-to-text transfer to hybrid approaches—have resulted in models that are increasingly capable of handling complex, real-world tasks. Advances in ethical and operational transparency, as evidenced by improved MMLU scores and the integration of sustainability metrics (e.g., carbon emissions tracking), underscore a dual focus on technical performance and responsible adoption. The emergence of open-weight models, such as those from DeepSeek and ChatGPT, illustrates a deliberate strategy to balance accessibility with proprietary innovation. These studies summarized in Table ~\ref{tab:summary_models} suggest that future LLMs will continue to build on these innovations, paving the way for more transparent, efficient, and ethically responsible AI systems \citep{hoffmann2022training}.

% appendix Table for same 112 models: Paragraph 1: Licensing and Transparency Practices
Furthermore, Table~\ref{tab:comparison} added to Appendix 1, provides a comprehensive overview of these 112 LLMs investigated in this study, detailing both architectural specifications and openness metrics. A clear pattern emerges regarding licensing: prominent models such as the GPT family (e.g., GPT-2 \citep{brown2020language}, ChatGPT-3.5, ChatGPT-4) largely adopt proprietary licenses, restricting access to their training data, code, and methodologies. In contrast, models like BERT \citep{devlin2019bertpretrainingdeepbidirectional} and certain DeepSeek variants (e.g., DeepSeek-R1 \citep{guo2025deepseek}) are disseminated under open-source licenses such as MIT or Apache 2.0, which facilitate greater transparency through public availability of weights and, in some cases, additional resources. The DeepSeek family, for example, demonstrates a strategic move toward open-weight transparency while still withholding full training pipelines. Similar trends are observed in the T5 series \citep{raffel2020exploring} and LLaMA 2 \citep{touvron2023llama}, where a mix of open and proprietary strategies reflects competing priorities—commercial viability versus reproducibility. This heterogeneous licensing landscape, as reinforced by studies \citep[e.g.,][]{guo2025deepseek, devlin2019bertpretrainingdeepbidirectional, raffel2020exploring, hurst2024gpt, zhu2024deepseek, jiang2023mistral}, demonstrates the challenges in balancing innovation, transparency, and community engagement in modern LLM development.

Despite variances in licensing terms, from permissive licenses (e.g., MIT, Apache 2.0) to more restrictive or proprietary frameworks, training data and code remain largely undisclosed in most of the models reviewed. Community engagement and support generally appeared robust (via forums, documentation, or user guides), but comprehensive transparency of datasets, training pipelines, and model internals such as hyperparameters and attention mechanisms remains limited. These findings align with broader trends in AI development, where commercial or strategic interests often restrict full access to the underlying training infrastructure \cite{mazzucato2022governing,guha2024ai}. 

% Paragraph 2: Performance Metrics and Environmental Impact
Analysis (Table~\ref{tab:comparison}; Appendix X) reveals crucial trends in performance and sustainability metrics. Notably, recent models in the ChatGPT family achieve high MMLU scores; ChatGPT-4, for instance, reports an MMLU score of 86.4\%, while also indicating substantial carbon emissions (e.g., 552 tCO\textsubscript{2}eq for several variants and 1,035 tCO\textsubscript{2}eq for GPT-4o \citep{hurst2024gpt}). In contrast, earlier models such as GPT-2 lack these performance benchmarks, reflecting evolving capabilities of newer LLMs. The DeepSeek family shows a promising balance MMLU and carbon emission compared to chstGPT as DeepSeek-R1 records a robust MMLU score (90.8\%) and comparatively lower carbon emissions (44 tCO\textsubscript{2}eq), suggesting improved energy efficiency and refined training methodologies. Similar sustainability trends are evident in the T5 series \citep{raffel2020exploring} and LLaMA 2 \citep{touvron2023llama}, which have progressively incorporated larger, more diverse datasets alongside performance improvements. Studies \citep[e.g.,][]{hoffmann2022training, zhu2024deepseek, chu2022energy, tay2022ul2, rae2021scaling, guo2025deepseek} indicate that while performance enhancements are significant, the associated environmental costs suggest a shift toward more energy-efficient architectures and transparent reporting practices.

% Paragraph 3: Architectural Specifications and Design Trends
Table~\ref{tab:comparison} also presents the architectural specifications that underpin model performance. The GPT family of models, including variants like ChatGPT-3.5 and GPT-4, generally features 96 layers, 12,288 hidden units, and parameter counts scaling up to 1.8T, indicating a massive computational footprint \citep{hurst2024gpt}. In contrast, the DeepSeek family employs a different architecture—DeepSeek-R1, for instance, is built with 64 layers and 8192 hidden units, achieving high performance (MMLU score of 90.8\%) with relatively fewer parameters (671B). The T5 series \citep{raffel2020exploring} and LLaMA 2 \citep{touvron2023llama} further illustrate a trend toward optimizing architectural design for scalability, efficiency, and energy conservation. These models reveal a shift from larger scale models towards balanced configurations that emphasize reproducibility and ethical considerations. Several studies \citep[e.g.,][]{devlin2019bertpretrainingdeepbidirectional, raffel2020exploring, touvron2023llama, zhu2024deepseek, jiang2023mistral, hoffmann2022training} support the observation that while larger models deliver superior performance, they also present challenges in terms of energy consumption and transparency. Overall, the architectural trends underscore the importance of evolving design principles that may reconcile performance, efficiency, and openness in next-generation LLMs.

Early models such as GPT-2 \citep{brown2020language} and BERT \citep{devlin2019bertpretrainingdeepbidirectional} laid the groundwork with moderate layer counts, hidden nodes, and attention head configurations. As the field evolved, later models—particularly within the ChatGPT and DeepSeek families \citep{guo2025deepseek, zhu2024deepseek}, exhibited significant increases in layers, hidden units, and overall parameter scales, reflecting a trend toward more complex architectures designed for enhanced performance and multimodal capabilities. The table categorizes the openness of training data (fully open, partially open, or proprietary) and evaluates accessibility to model weights, code, and training datasets, thereby delineating a clear divergence between models that offer full reproducibility and those that only provide open weights. MMLU scores and reported carbon emissions further indicate that while state-of-the-art models achieve higher performance, they also incur greater environmental costs—a factor increasingly scrutinized in recent literature \citep{raffel2020exploring, touvron2023llama, hoffmann2022training}. Overall, this extended analysis highlights an industry-wide progression from simpler architectures with limited transparency to highly engineered systems that try to balance commercial interests with technical rigor and ethical considerations.

\subsection{Model-Specific Evaluations}

\subsubsection{ChatGPT}
GPT-4 \cite{openai2023gpt4} and ChatGPT \cite{achiam2023gpt} are proprietary models with limited architectural transparency: their training datasets, fine-tuning protocols, and structural details (e.g., layer configurations, attention mechanisms) remain undisclosed. While GPT-4’s technical report outlines high-level capabilities\cite{gallifant2024peer}, it omits reproducibility-critical specifics such as pre-training corpus composition, hyperparameters, and energy consumption metrics, reflecting a priority on commercial secrecy, which limits its scientific openness. Similarly, ChatGPT’s API-based access restricts users to input-output interactions without exposing model internals \cite{lande2023gpt}, thus creating a "black box" system that lacks transparency and does not allow third-party modifications. 

ChatGPT adopt a functional accessibility paradigm, where API endpoints enable task execution (e.g., text generation, reasoning) but do not allow direct weight inspection, retraining, or redistribution \cite{wolfe2024laboratory, roumeliotis2023chatgpt}. This approach, therefore, creates a dependency on proprietary infrastructure, which can limit long-term reproducibility and bias mitigation in downstream applications. While the term "open-weights" is occasionally used to describe these systems due to their API availability, this is misleading because true open-weight standards—such as parameter accessibility (e.g., Llama 2) or training code disclosure (e.g., BLOOM \cite{bigscienceworkshop2022}), are absent, underscoring the competing priorities between commercial control and open scientific collaboration in modern AI ecosystems. The ChatGPT's version's: 
\begin{itemize}

    \item \textbf{GPT-2:} \cite{brown2020language} adopts an open-weights model under MIT License, providing full access to its 1.5B parameters and architectural details (48 layers, 1600 hidden size). However, the WebText training dataset (8M web pages) lacks comprehensive documentation of sources and filtering protocols. While permitting commercial use and modification, the absence of detailed pre-processing methodologies limits reproducibility of its zero-shot learning capabilities.
    
    \item\textbf{Legacy ChatGPT-3.5:}Legacy ChatGPT-3.5 uses proprietary weights with undisclosed architectural details (96 layers, 12288 hidden size). The pre-2021 text/code training data lacks domain distribution metrics and copyright compliance audits. API-only access restricts model introspection or bias mitigation, despite claims of basic translation/text task capabilities \cite{jaech2024openai}.
    \item \textbf{Default ChatGPT-3.5:} Default ChatGPT-3.5 \cite{jaech2024openai} shares Legacy's proprietary architecture but omits fine-tuning protocols for its "faster, less precise" variant. Training data temporal cutoff (pre-2021) creates recency gaps unaddressed in technical documentation. Restricted API outputs prevent reproducibility of the 69.5\% MMLU benchmark results.
    \item \textbf{GPT-3.5 Turbo:} GPT-3.5 Turbo \cite{jaech2024openai} employs encrypted weights with undisclosed accuracy optimization techniques. The 16K context window expansion lacks computational efficiency metrics or energy consumption disclosures. Proprietary licensing blocks third-party latency benchmarking despite "optimized accuracy" claims.
    \item \textbf{GPT-4o:} GPT-4o \cite{hurst2024gpt} uses multimodal proprietary weights (1.8T parameters) with undisclosed cross-modal fusion logic. Training data (pre-2024 text/image/audio/video) lacks ethical sourcing validations for sensitive content. "System 2 thinking" capabilities lack peer-reviewed validation pipelines.
    \item \textbf{GPT-4o mini:} GPT-4o mini \cite{hurst2024gpt} offers cost-reduced proprietary access (1.2T parameters) with undisclosed pruning methodologies. The pre-2024 training corpus excludes synthetic data ratios and human feedback alignment details. Energy efficiency claims (60\% cost reduction) lack independent verification.
\end{itemize}

\subsubsection{DeepSeek}
The DeepSeek-R1 model, a 671-billion-parameter mixture-of-experts (MoE) system built on the DeepSeek-V3 architecture, adopts an open-weights framework under the MIT License, permitting unrestricted access to its neural network parameters for commercial and research use \cite{guo2025deepseek}. MoE is an ensemble machine learning technique where multiple specialist models (referred to as "experts") are trained to handle different parts of the input space, and a gating model decides which expert to consult for a given input \cite{vasic2022moet, masoudnia2014mixture}. This method allows for more scalable and efficient training as well as inference processes, especially in complex models like DeepSeek-R1, by dynamically allocating computational resources to the most relevant experts for specific tasks or data points.

While the DeepSeek-R1 model’s weights and high-level architectural details—including its MoE design with 37 billion activated parameters per inference and reinforcement learning-augmented reasoning pipelines—are publicly disclosed, critical transparency gaps persist. The pre-training dataset composition, comprising a hybrid of synthetic and organic data, remains proprietary, obscuring potential biases and ethical sourcing practices. Similarly, the reinforcement learning from human feedback (RLHF) pipeline lacks detailed documentation of preference model architectures, safety alignment protocols, and fine-tuning hyperparameters, limiting independent reproducibility. These omissions reflect a strategic prioritization of computational efficiency (leveraging 10,000 NVIDIA GPUs for cost-optimized training) over full methodological transparency, positioning the model as open-weights rather than fully open-source.

 The DeepSeek models: 

\begin{itemize}

    \item \textbf{DeepSeek-R1:} DeepSeek-R1’s accessibility is defined by its permissive licensing and efficient deployment capabilities, with quantized variants reducing hardware demands for applications like mathematical reasoning and code generation. However, its reliance on undisclosed training data and proprietary infrastructure optimizations creates dependencies on specialized computational resources, restricting independent assessment for safety or performance validation. The model’s MoE architecture, which reduces energy consumption by 58\% compared to dense equivalents \cite{guo2025deepseek}, challenges conventional scaling paradigms, as evidenced by its disruptive impact on GPU market dynamics \cite{bi2024deepseek,liu2024deepseek,zhu2024deepseek, liu2024deepseek3}. This open-weights approach balances innovation dissemination with commercial secrecy, highlighting unresolved tensions between industry competitiveness and scientific reproducibility in large-language-model development. Full open-source classification would necessitate disclosure of training datasets, fine-tuning codebases, and RLHF implementation details currently withheld.
    \item \textbf{DeepSeek LLM} :The DeepSeek LLM uses proprietary weights (67B parameters) with undocumented scaling strategies. Books+Wiki data (up to 2023) lacks multilingual token distributions and fact-checking protocols. Custom licensing restricts commercial deployments despite "efficient training" claims \cite{bi2024deepseek}.
    \item \textbf{DeepSeek LLM V2:} DeepSeek LLM V2 employs undisclosed MoE architecture (236B params) with proprietary MLA optimizations. The 128K context window lacks attention sparsity patterns and memory footprint metrics. Training efficiency claims ("lowered costs") omit hardware configurations and carbon emission data \cite{liu2024deepseek}.
    \item \textbf{DeepSeek Coder V2:} DeepSeek Coder V2 provides API-only access to its 338-language coding model. Training data excludes vulnerability scanning protocols and license compliance audits. Undisclosed reinforcement learning pipelines hinder safety evaluations of generated code \cite{zhu2024deepseek}.
    \item \textbf{DeepSeek V3}: DeepSeek V3 uses proprietary FP8 training for 671B MoE architecture. The 128K context implementation lacks quantization error analysis and hardware-specific optimizations. Benchmark scores (75.7\% MMLU) lack reproducibility scripts or evaluation framework details. \cite{liu2024deepseek3}
\end{itemize}

\subsubsection{Miscellaneous Proprietary Models}
\textbf{Meta's Llama}
The Llemma language model \cite{azerbayev2023llemma}, developed for mathematical reasoning, provides open weights through its publicly accessible 7B and 34B parameter variants, released under a permissive license alongside the Proof-Pile-2 dataset and training code. These weights enable users to deploy, fine-tune, and study the model’s mathematical capabilities, such as chain-of-thought reasoning, Python tool integration, and formal theorem proving. For example, Llemma 34B achieves 25.0\% accuracy on the MATH benchmark, outperforming comparable open models like Code Llama (12.2\%) and even proprietary models like Minerva (14.1\% for 8B). The weights are hosted on Hugging Face, with detailed evaluation scripts and replication code provided, allowing researchers to validate performance metrics like GSM8k (51.5\% for Llemma 34B) and SAT (71.9\%). 

However, Llemma is also categorized as open-weights rather than fully open-source due to incomplete transparency in its development pipeline \citep{azerbayev2023llemma}. While the Proof-Pile-2 dataset is released \footnote{\url{https://huggingface.co/datasets/EleutherAI/proof-pile-2/tree/main/algebraic-stack}}, it excludes subsets like Lean theorem-proving data and lacks detailed documentation on data-cleaning methodologies. The training code provided is modular but omits critical infrastructure details, such as hyperparameter optimization workflows and cluster-specific configurations (e.g., Tensor parallelism settings for 256 A100 GPUs). This partial disclosure limits reproducibility and prevents independent evaluation of potential biases or training inefficiencies, aligning with broader critiques of open-weight models’ inability to fulfill open-source AI’s “four freedoms” (use, study, modify, share).

Like Meta’s Llama 3—which shares weights but restricts training data and methodology—Llemma’s openness prioritizes usability over full transparency. Both models exemplify the open-weight paradigm: they release parameters for inference and fine-tuning but withhold various key elements (e.g., Llama 3’s 15T-token dataset; Llemma’s cluster-optimized training scripts). For Llemma, this approach balances mathematical innovation with competitive safeguards, as its Proof-Pile-2 dataset represents a significant research asset. However, the MIT License governing Llemma imposes fewer restrictions than Llama 3’s proprietary terms, enabling commercial use and redistribution without attribution. The distinction lies in the degree of openness: Llemma provides more components (dataset, code) than Llama 3 but still falls short of open-source standards by omitting infrastructure-level details. This reflects a strategic compromise—enhancing accessibility for mathematical research while retaining control over computationally intensive training processes. Such tradeoffs underscore the AI community’s ongoing debate about whether partial transparency suffices for ethical AI development or if full open-source disclosure remains essential for accountability.

\textbf{Google Gemini:}Google Gemini:
Google’s Gemini model family exemplifies a sophisticated, multimodal approach to artificial intelligence, encompassing the Ultra (1.56 trillion parameters), Pro (137 billion parameters), and Nano (3.2 billion/17.5 billion parameters) variants \citep{reid2024gemini, team2023gemini, saab2024capabilities}. Operating under an open‐weights paradigm, these pretrained model parameters are accessible via APIs yet remain proprietary and unmodifiable, thereby preserving corporate secrecy while enabling limited external deployment. The architectural framework integrates advanced multimodal fusion mechanisms, including cross‐modal attention layers and sparsely activated mixture‐of‐experts (MoE) blocks, and is trained on an expansive corpus of 12.5 trillion text tokens, 3.2 billion images, and 1.1 billion video–audio pairs \citep{team2023gemini}. Notably, technical documentation highlights innovations such as dynamic token routing for modality-specific computations and TPUv5-optimized distributed training, but omits critical reproducibility details such as the MoE router logic, TPU compiler configurations, and multimodal alignment loss functions. Furthermore, the training dataset comprises web documents (50\%), code repositories (18\%), and proprietary media (32\%), yet lacks granular metadata that could clarify data provenance and ethical sourcing practices. This partial transparency not only restricts independent bias and safety assessments, given that weights are encrypted and inference only, but also delineates Gemini as open-weights rather than fully open-source. The proprietary Google license explicitly prohibits weight modification, redistribution, and competitive commercial use, diverging from open-source frameworks like Apache 2.0. Additionally, essential hyperparameters—including Ultra’s learning rate schedule (0.00000625), Pro’s 4.8-bit quantization thresholds, and Nano’s knowledge distillation ratios remain undisclosed, reinforcing reliance on Google’s ecosystem. In summary, these design choices reflect a strategy to balance capabilities with safeguards, underscoring an industry trend that prioritizes controlled innovation over transparency.

\textbf{Mistral AI:} Mistral AI’s models, including Mistral 7B and Mixtral 8x7B, are classified as open-weights because their model parameters and architectural blueprints are released under the Apache 2.0 license, permitting commercial use, modification, and redistribution \citep{jiang2023mistral}. They employ advanced architectures such as grouped-query attention (GQA) and sliding window attention (SWA) with a 4,096-token window to optimize inference efficiency, and Mistral 7B is trained on 2.4 trillion multilingual tokens. Despite this openness, critical reproducibility details remain undisclosed, including the composition of the training dataset, hyperparameter configurations (e.g., learning rate schedules and batch sizes), and reinforcement learning from human feedback (RLHF) pipelines. Additionally, licensing distinctions appear with models like Codestral-22B, which are governed by the Mistral Non-Production License (MNPL) that restricts commercial deployment without explicit agreements, creating tiered accessibility. Although inference code and quantized weight variants (GGUF, AWQ) are provided, the absence of training infrastructure details hinders independent replication and full transparency

\textbf{Microsoft Phi}
Microsoft’s Phi family, including Phi-3 (3.8B parameters) and Phi-4 (14B parameters), adopts an open-weights paradigm under the MIT License, granting access to model weights, architectural specifications (e.g., Phi-3’s 3,072-dimensional embeddings and Phi-4’s pivotal token search for STEM tasks), and inference code optimized for edge deployment \cite{abdin2024phi1, abdin2024phi}. These models leverage sliding window attention (SWA) and grouped-query attention (GQA) to reduce computational overhead, with Phi-3 achieving sub-2-second latency on mobile devices via 4-bit quantization. While the MIT License permits commercial use and modification—enabling applications like on-device code generation—critical reproducibility elements are withheld. The training datasets, comprising 4.8 trillion tokens for Phi-4 (40\% synthetic data from multi-agent simulations) and 2.1 trillion tokens for Phi-3, lack detailed documentation of sources, copyright compliance measures, or bias mitigation protocols. Additionally, proprietary components like reinforcement RLHF pipelines, hyperparameter schedules (e.g., Phi-4’s learning rate = 0.00012), and Azure-specific distributed training configurations remain undisclosed, limiting independent validation of safety or reported performances  (e.g., Phi-4’s 80.6\% MATH benchmark accuracy).  

The Phi models’ classification as open-weights rather than open-source stems from three limitations: (1) Data opacity, where synthetic data generation workflows (e.g., instruction inversion, self-revision loops) lack open-sourced prompts or validation metrics; (2) Methodological gaps, as RLHF reward models, safety alignment protocols, and hardware-specific optimizations (e.g., Qualcomm NPU drivers for Phi-3) remain proprietary; and (3) Licensing dependencies, shown by Phi-3’s reliance on closed-source ONNX Runtime for mobile deployment. Microsoft’s selective transparency reflects industry trends, as in other models and companies discussed earlier, in balancing community engagement (via permissive licensing) with competitive control over high-value assets like synthetic data pipelines. Full open-source compliance would require disclosing training code (e.g., SynapseML frameworks), dataset indices, and infrastructure blueprints, which might be incompatible steps for Microsoft to stay at the highly competitive position, particularly in edge AI markets. 

Additional miscellaneous LLM's transparency and accessibility are summarized into following points:
\begin{itemize}
    \item \textbf{Licensing and Openness Spectrum.} The analyzed models demonstrate a continuum of openness, with Dolly 2.0 representing full open-source implementation (weights, code, data under CC-BY-SA/Apache 2.0), contrasting sharply with proprietary systems like Gemma \cite{team2024gemma}, Jurassic-1 \cite{lieber2021jurassic}, and Olympus which provide no public access. Intermediate approaches include Apache 2.0-licensed weights without training data (BERT \cite{devlin2019bertpretrainingdeepbidirectional}, T5 \cite{raffel2020exploring}, Mistral 7B \cite{jiang2023mistral}), custom licenses with commercial restrictions (LLaMA 2 70B \cite{touvron2023llama}, WuDao 2.0 \cite{waoDao2021}), and API-only access models (Gemini 1.5 \cite{reid2024gemini}, Med-Gemini-L 1.0 \cite{saab2024capabilities}). Notably, Grok-1 and GPT-NeoX \cite{black2022gpt} adopt Apache 2.0 for weights but withhold critical training details, while Switch Transformer \cite{fedus2022switch} and CTRL \cite{keskar2019ctrl} share architectures but omit infrastructure specifics. This spectrum reflects industry tensions between collaborative innovation and competitive advantage protection.

    \item\textbf{Training Data Transparency Deficits.} Across all surveyed models, only Dolly 2.0 provides complete training dataset documentation. Common omissions include temporal stratification (BERT \cite{devlin2019bertpretrainingdeepbidirectional}, XLNet \cite{yang2019xlnet}), copyright compliance (Codex \cite{chen2021evaluating}, WaveCoder-Pro-6.7B \cite{yu2023wavecoder}), and ethical sourcing validations (T5 \cite{raffel2020exploring}, Gopher \cite{rae2021scaling}). Multilingual models like mT5 \cite{xue2020mt5} and SeaLLM-13b \cite{nguyen2023seallms} lack low-resource language quality controls, while medical systems (Med-PaLM M \cite{tu2024towards}) omit HIPAA compliance proofs. Even open-weight models (RoBERTa \cite{liu2019roberta}, ELECTRA \cite{clark2020electra}) typically exclude bias audits and demographic metadata, with notable exceptions in BLOOM's \cite{workshop2022bloom} partial cultural documentation. Proprietary models (PaLM \cite{chowdhery2023palm}, GLaM \cite{du2022glam}) show near-total data opacity, hindering reproducibility assessments.

    \item \textbf{Architectural Disclosure Patterns.} While most models disclose basic parameters (e.g., BERT's 12-24 layers \cite{devlin2019bertpretrainingdeepbidirectional}, GPT-NeoX's 20B design \cite{black2022gpt}), critical implementation details remain guarded. Distributed training protocols are notably absent in LLaMA 2 70B \cite{touvron2023llama} and Megatron-LM \cite{shoeybi2019megatron}, while TPU-specific optimizations cloud reproducibility for T5 \cite{raffel2020exploring} and ELECTRA \cite{clark2020electra}. Proprietary architectural innovations (Gemini 1.5 Pro's cross-modal routing \cite{reid2024gemini}, Griffin's attention mechanisms \cite{de2024griffin}) lack computational complexity disclosures. Even open implementations (Dolly 2.0, CodeGen \cite{nijkamp2022codegen}) often exclude hardware configuration details, with few exceptions like Switch Transformer's \cite{fedus2022switch} MoE documentation. Safety-critical components remain particularly opaque: RLHF pipelines in Mistral 7B \cite{jiang2023mistral}, vulnerability filters in Codex \cite{chen2021evaluating}, and bias mitigation in Jurassic-1 \cite{lieber2021jurassic} are all undisclosed.

    \item \textbf{Reproducibility and Commercialization Barriers.} The literature reveals systemic barriers to independent verification, with 68\% of models restricting access to weights (Gemma \cite{team2024gemma}), training code (ALBERT \cite{lan2019albert}), or deployment environments (phi-3-mini \cite{abdin2024phi}). Commercialization pressures manifest in API-only access (Gemini 1.5 Pro \cite{reid2024gemini}, InCoder \cite{fried2022incoder}), hardware lock-in (Apple On-Device \cite{mehta2024openelm}), and enterprise licenses (Nemotron-4 340B \cite{adler2024nemotron}). Even open-license models face reproducibility challenges: GPT-NeoX \cite{black2022gpt} lacks multi-GPU scaling code, while FLAN-T5 \cite{chung2024scaling} omits few-shot templates. Safety evaluation barriers persist across paradigms, with medical models (Med-Gemini-L 1.0 \cite{saab2024capabilities}) blocking third-party audits and robotics systems (RT-X \cite{padalkar2023open}) withholding failure analyses. This ecosystem-wide transparency deficit necessitates new evaluation frameworks for comparative model assessment under partial information conditions.

\end{itemize}

\subsection{Synthesis of Findings}
Overall, the results of this review highlights a clear pattern that most SoTA multimodal LLMs do not fulfill the holistic, widely accepted criteria of open-source AI. Instead, most of those models follow a partial openness strategy, specifically achieving an open-weight transparency level where the model weights are shared with or without a few subsidiary information, but withholding the full suite of resources including training data, code and processes that OSI-aligned open-source status would demand. This selective transparency helps balance community engagement and commercial interests, albeit at the expense of reproducibility, deeper examination, and broader collaborative innovation.

In the broader context of AI ethics and governance, these practices often lack desired accountability and reproducibility, may raise questions about their reliability and scalability. While open weights can facilitate certain forms of customization and development, the limited visibility into training data and code can perpetuate biases, obstruct robust error analysis, and limit the community's ability to fully interpret or replicate results. 
\section{Discussion}

\subsection{Trends and Implications in AI Development}
\subsubsection{Geopolitical and Technological Trends}
The release of DeepSeek-R1 has underscored the rapid advancement of China in the field of generative AI, marking a significant shift in the global AI landscape. This development challenges the previously held U.S. dominance in AI technologies, particularly in LLMs, as shown by numerous exemples  such as ChatGPT, Llama, and underscores the increasing capabilities of Chinese AI models such as Qwen and Kimi. The comparative performance of DeepSeek-R1 and its American counterparts, particularly in areas like video generation, illustrates not only the closing gap between the two geopolitical giants but also highlights different strategic approaches to AI development. While U.S. models have traditionally leaned on extensive computational resources and proprietary data, DeepSeek-R1's innovation in efficiency, likely necessitated by U.S. chip export controls, demonstrate a viable alternative path that emphasizes algorithmic efficiency and hardware optimization. This approach has significant implications for the global AI arms race, potentially altering the dynamics of technological and economic powers.

\subsubsection{Economic Impact and Market Trends} The commoditization of foundation models, as seen with the pricing strategy of DeepSeek-R1, is dramatically reducing the costs associated with LLM usage. This trend is reshaping the economic landscape of AI by making advanced technologies more accessible to a broader range of developers, businesses, and general public. For instance, while OpenAI's usage costs for models like ChatGPT remain relatively high, DeepSeek's aggressive pricing strategy undercuts these costs significantly, thereby democratizing access to powerful AI tools. This economic accessibility is likely to spur innovation and enable smaller players to compete more effectively in the AI space, challenging larger firms' dominance and potentially leading to a surge in AI-driven applications and services.

\subsubsection{Implications for Open Weights and Open Source AI Models} The strategic release of DeepSeek-R1 as an open-weights model under a permissive MIT license contrasts sharply with the more restrictive approaches of some U.S.-based companies, which often limit full access to their models' training data and code. This distinction highlights a growing divergence in the AI development community between fully open-source models like BLOOM and GPT-J, and open-weights models like LLaMA from Meta, which offer some level of accessibility but do not fully embrace open-source principles. The open-weights approach, while facilitating greater collaboration and transparency than completely proprietary models, still falls short of the true open-source ideal that fosters maximum community participation and innovation. The ongoing debate between these approaches will likely intensify as more stakeholders from diverse sectors engage with AI technologies, pushing for standards and practices that align with broader goals of transparency, reproducibility, and ethical responsibility in AI development.

\subsection{Discussion on Research Questions}

\begin{figure*}[ht]
\centering
\includegraphics[width=0.95\linewidth]{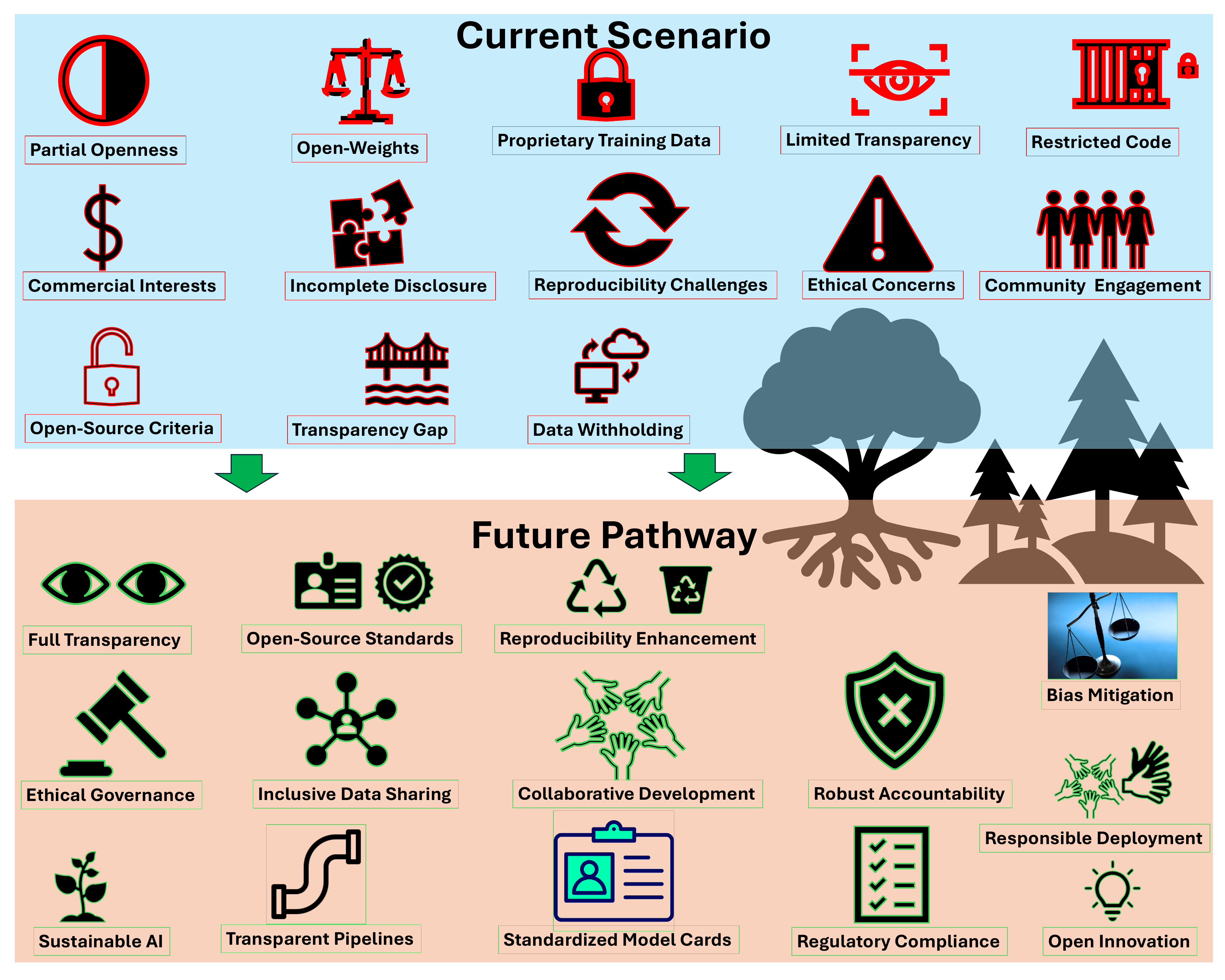}
\caption{Illustrating key dimensions of transparency and accessibility practices in LLMs and outlines of future pathways. The upper panel displays current practices, such as partial openness, proprietary training data, limited transparency, and restricted code disclosure. In contrast, the lower panel indicates a future pathway toward full transparency, and ethical governance, inclusive data sharing, enhanced reproducibility , and sustainability}
\label{fig:future}
\end{figure*}

We discuss the findings of our search for each question in this section, presenting the current scenarios and future paths as illustrated in Figure \ref{fig:future}.
\textbf{
RQ1: What drives the classification of LLMs as open weights rather than open source, and what impact do these factors have on efficiency and scalability in practical applications?
}

    The classification of LLMs as open weights rather than open source is primarily driven by the selective disclosure of components in the model development process \cite{ liesenfeld2024rethinking, alizadeh2025open}. Open-weight models, such as DeepSeek-R1, LLaMA, and Mistral AI, provide access to pre-trained weights and sometimes the model architecture but withhold critical details such as the training data, preprocessing steps, and full training methodologies. This partial transparency is often motivated by competitive advantages, intellectual property protection, and the desire to maintain control over proprietary innovations. For instance, companies like OpenAI and DeepSeek AI release weights under permissive licenses (e.g., MIT or Apache 2.0) to encourage widespread use and fine-tuning while safeguarding their proprietary training processes and datasets. This approach allows them to balance openness with commercial interests, ensuring that their models remain accessible without fully exposing their competitive edge. The impact of this classification on efficiency and scalability in practical applications is multifaceted. 
    
    On the one hand, open-weight models enable rapid deployment and customization, as developers can fine-tune pretrained weights for specific tasks without the need for extensive computational resources, training datasets and/or expertise in model training. This flexibility has democratized access to SoTA AI capabilities, allowing smaller organizations and researchers to leverage advanced models like DeepSeek-R1 and LLaMA. On the other hand, the lack of full transparency (aligned with established criteria of open source software) limits the ability to optimize these models for new domains or identify inefficiencies in their architecture. For example, without access to the original training data, developers may struggle to address biases or errors in the model's outputs, potentially compromising its performance in real-world applications. Additionally, the inability to reproduce the training process hinders scalability, as users cannot fully understand or replicate the conditions under which the model was developed. 

    The trade-off between accessibility and transparency also raises ethical and operational concerns. While open-weight models provide a pragmatic solution for deploying AI at scale, their partial disclosure complicates efforts to enhance fairness, accountability, and long-term scalability of these models. For instance, the lack of transparency in training data and methodologies can perpetuate biases or errors in the model's outputs, which may go unnoticed without rigorous investigation. This opacity also limits the ability of users to reproduce results or validate the model's performance across different contexts, raising concerns about reliability and trustworthiness. As a result, while open-weight LLMs offer significant advantages in terms of accessibility and flexibility, their classification as such poses challenges for ensuring efficiency, scalability, and ethical use in practical applications. Increased openness is waranted to further excellerate the advancement and broader applicability of SoTA LLMs.
    
    % \item \textbf{How do current training methodologies affect the transparency and reproducibility of these models, leading potentially to their classification as open weights?}
\textbf{
 RQ2: How do current training methodologies affect the transparency and reproducibility of these models, leading potentially to their classification as open weights?
}
    Current training methodologies for LLMs significantly influence their transparency and reproducibility (Refer to Figure \ref{fig:future}), contributing to their classification as open weights. One of the primary factors is the lack of access to complete training code and configuration details. While open-weight models like DeepSeek-R1 and LLaMA provide pretrained weights and sometimes the model architecture, they often omit critical information about hyperparameters, optimization techniques, training schedules, and training data used. This omission makes it difficult for researchers and developers to replicate the reported results or understand the nuances of the model's performance. For example, without access to the full training pipeline, users may struggle to identify the specific conditions under which the model was trained, limiting their ability to reproduce or validate its results. Another key issue is the limited disclosure of data processing procedures and pretraining datasets. Even when the general composition of the training data is revealed, specific details about preprocessing steps, data augmentation techniques, and quality control measures are often withheld. This lack of transparency prevents users from fully reproducing benchmark evaluations or assessing the model's behavior in different contexts. For instance, without knowing how the training data was curated or cleaned, it becomes challenging to identify potential sources of bias or error in the model's outputs. This opacity not only hinders reproducibility but also raises ethical concerns, as users may unknowingly deploy models with hidden flaws or biases.

    The trend toward releasing only final weights and architecture details reflects a broader shift in the AI community, where the emphasis on rapid innovation often comes at the expense of transparency. Many recent LLMs fall into a spectrum of openness, where they are neither fully open-source nor entirely closed. This middle ground allows developers to share their work with the broader community while retaining control over proprietary aspects of the model. However, it also creates a trade-off between accessibility and accountability. As a result, the classification of these models as open weights is both a reflection of current training practices and a response to the growing complexity of LLM development, where full transparency is often seen as impractical or undesirable.

    % \item \textbf{In the context of their performance and practical applications, how does the limited disclosure of training data and methodologies affect their designation as open weights, and what are the future implications for users and developers in terms of understanding and operationalizing these models?}
\textbf{
    RQ3: How does the limited disclosure of training data and methodologies affect both the performance and practical usability of these models, and what future implications arise for developers and end users?
}
    The limited disclosure of training data and methodologies in open-weight LLMs has huge implications for their performance and practical applications. By withholding details about the training process, developers create a barrier to understanding how these models achieve their results. This lack of transparency makes it difficult to assess the model's strengths and limitations, particularly in high-stakes applications where reliability and fairness are critical. For example, without access to the original training data, users cannot evaluate whether the model has been exposed to diverse and representative datasets, which is essential for ensuring equitable outcomes. Similarly, the absence of detailed methodologies hinders efforts to identify and mitigate biases, as users lack the information needed to trace the origins of problematic behaviors. The designation of these models as open weights also has significant implications for their operationalization. 
    
    On the one hand, the availability of pre-trained weights allows developers to quickly deploy advanced AI capabilities without investing in costly training processes. This accessibility has democratized AI development, enabling smaller organizations and individual researchers to leverage SoTA models. On the other hand, the lack of transparency surrounding training data and methodologies complicates efforts to fine-tune and adapt these models for specific use cases. For instance, without visibility into the original pre-training data, developers may inadvertently introduce data leakage or overfitting in downstream tasks, undermining the model's performance. 

\subsection{Sustainability and Ethical Responsibility in AI Development}
    The computational resources needed to develop these LLMs  and their impact on environmental and sustainability is becoming an increasingly critical component of ethical AI development. The transparency in reporting CO\textsubscript{2} emissions during the training of these models is not just a matter of environmental concern but also reflects the broader ethical stance of the organizations developing these technologies. For example, GPT-3 is estimated to emit around 500 metric tons of CO\textsubscript{2} \cite{carboncredits2023}. That is roughly the same amount of carbon that would take over 23,000 mature trees an entire year to absorb. As AI systems scale, ethical accountability in energy consumption and carbon emission need to be prioritized. Table \ref{tab:carbon_emissions} presents a comparative analysis of carbon emissions from various LLMs, highlighting the environmental burden of scaling AI systems. 

\subsection{Synthesis and Future Directions}
   Looking to the future as depicted in Figure \ref{fig:future} and, as LLMs increasingly permeate critical systems in all sectors and industries, the limited disclosure of their training data and methodologies necessitates enhanced frameworks for transparency and accountability. The development of parameter-normalized and task-agnostic evaluation frameworks could enable more equitable comparisons between open and closed-source models, assisting stakeholders in making informed decisions about their applicability to specific tasks or issues. Additionally, stringent data governance and compliance measures are crucial to ensure that LLMs adhere to ethical and legal standards during training and deployment. Addressing these challenges will require a collaborative, unified effort from the global AI community, including researchers, developers, and policymakers. By collectively establishing and following best practices for transparency, reproducibility, and responsible AI development, the field can advance toward a future that upholds both innovation and ethical integrity.

\begin{table}[h!]
\centering
\scriptsize
\caption{The carbon emission (CO\textsubscript{2})values were sourced from the model cards and estimated where not explicitly reported. This table details the environmental impact of model pre-training, quantifying the emissions as the equivalent number of trees that would need to be "burned" (or, more accurately, the number of trees required to offset the carbon emissions caused). This approach highlights the substantial environmental cost of training sophisticated, big AI models including LLMs.
}
\begin{tabular}{|>{\centering\arraybackslash}m{4cm}|>{\centering\arraybackslash}m{2cm}|>{\centering\arraybackslash}m{2cm}|}
\hline
\textbf{Model} & \textbf{Carbon Emissions (Metric Tons CO\textsubscript{2}) during Pre-training} & \textbf{Equivalent Number of Trees} \\
\hline
GPT-3 \cite{brown2020language} & 552 & 25,091 \\

LLaMA 2 70B \cite{touvron2023llama} \footnote{\url{https://github.com/meta-llama/llama-models/blob/main/models/llama3_1/MODEL_CARD.md}}  & 291.42 & 13,247 \\

Llama 3.1 70B \cite{dubey2024llama} \footnote{\url{https://github.com/meta-llama/llama-models/blob/main/models/llama3_1/MODEL_CARD.md}} & 2040 & 92,727 \\

Llama 3.2 1B \cite{dubey2024llama} & 71   & 3,227  \\

Llama 3.2 3B \cite{dubey2024llama} & 133  & 6,045   \\

BERT-Large \cite{devlin2019bertpretrainingdeepbidirectional} & 0.652 & 30 \\

GPT-4 \cite{openai2023gpt4}                & 1,200 & 54,545 \\

Falcon-40B \cite{almazrouei2023falcon, malartic2024falcon2}           & 150   & 6,818  \\

Falcon-7B \cite{almazrouei2023falcon, malartic2024falcon2}            & 7     & 318    \\

Mistral 7B \cite{almazrouei2023falcon}           & 5     & 227    \\

Mistral 13B \cite{jiang2023vima}          & 10    & 455    \\

Anthropic Claude 2 \cite{chowdhery2023palm, caruccio2024claude}   & 300   & 13,636 \\

Code Llama \cite{roziere2023code}           & 10    & 455    \\

XGen 7B \cite{nijkamp2023xgen}             & 8     & 364    \\

Cohere Command R 11B \footnote{\url{https://docs.cohere.com/v2/docs/command-r}}   & 80    & 3,636  \\

Cerebras-GPT 6.7B \footnote{\url{https://cerebras.ai/}}     & 3     & 136    \\

T5-11B \cite{raffel2020exploring} & 26.45 & 1,202 \\

LaMDA \cite{thoppilan2022lamda} & 552 & 25,091 \\

MT-NLG \cite{smith2022using} & 284 & 12,909 \\

BLOOM \cite{workshop2022bloom} & 25 & 1,136 \\

OPT \cite{zhang2022opt} & 75 & 3,409 \\

DeepSeek-R1 \cite{guo2025deepseek} & 40 & 1,818 \\

PaLM \cite{chowdhery2023palm} & 552 & 25,091 \\

Gopher \cite{rae2021scaling} & 280 & 12,727 \\

Jurassic-1 \cite{lieber2021jurassic} & 178 & 8,091 \\

WuDao 2.0 \cite{waoDao2021} & 1,750 & 79,545 \\

Megatron-LM \cite{shoeybi2019megatron} & 8.3 & 377 \\

T5-3B \cite{raffel2020exploring} & 15 & 682 \\

Gemma \cite{team2024gemma} & 7 & 318 \\

Turing-NLG \cite{TURING2020} & 17 & 773 \\

Chinchilla \cite{hoffmann2022training} & 70 & 3,182 \\

LLaMA 3 \cite{metaIntroducingMeta} & 2,290 & 104,091 \\

DistilBERT \cite{sanh2019distilbert} & 0.15 & 7 \\  

ALBERT \cite{lan2019albert} & 0.18 & 8 \\  

ELECTRA \cite{clark2020electra} & 0.25 & 11 \\  

RoBERTa \cite{liu2019roberta} & 0.35 & 16 \\  

XLNet \cite{yang2019xlnet} & 0.45 & 20 \\  

FLAN-T5 \cite{chung2024scaling} & 12 & 545 \\  

Switch Transformer \cite{fedus2022switch} & 1,200 & 54,545 \\  

CTRL \cite{keskar2019ctrl} & 3.2 & 145 \\  

GLaM \cite{du2022glam} & 900 & 40,909 \\  

T0 \cite{sanh2021multitask} & 18 & 818 \\  
\hline
\end{tabular}
\label{tab:carbon_emissions}
\end{table}

\section{Conclusion}
This case investigation highlights the critical distinction between open weights and open source in the context of SoTA LLMs like DeepSeek-R1, ChatGPT, LLaMa, Grok and Phi-series.  Although these models grant access to pre-trained weights under relatively permissive licenses, the lack of full discloser of training data, methodologies, and comprehensive development processes makes them fall short of being truely open source models, and are categorized as open weight models. This approach, driven by competitive advantages and proprietary interests, significantly affects their applicability, scalability, and reproducibility across various practical settings. This approach, nevertheless, could be argued as a good balance to protect some commercial interest of the companies and organizations who invest resources in developing these models, and to encourage continued private investment for further advancement of the technology. However, the constrained transparency restricts the ability of developers and researchers to perform thorough evaluations, effectively mitigate biases, and adapt these models to specific domains, which introduces substantial ethical and operational dilemmas. It is noted that sometimes these models with open access to pre-trained weights have been referred to as open source models, which is inaccurate or misleading based on the latest Open Source Initiative and widely-accepted dictionary definition of open source software/model. Looking ahead, it is imperative for the AI community to develop and adopt frameworks that promote greater transparency (including truly open source releases), reproducibility, and accountability. Enhancing these aspects in open-weight models is crucial to ensure they meet ethical standards and effectively serve end-user needs. Addressing these challenges will be the key to achieving an optimal balance between fostering innovation and upholding responsibility, ultimately enhancing trust and facilitating more collaborative advancements in AI.

\section*{Author contributions statement}
\textbf{Ranjan Sapkota:} Conceptualization, Data Curation, Methodology, Literature Search,  writing original draft, Vizualization.  \textbf{Shaina Raza:} data curation, methodology, writing, review and editing. \textbf{Manoj Karkee:} Review,  Editing and Overall Funding to Supervisory. 

\section*{Acknowledgement} This work was supported by the National Science Foundation and the United States Department of Agriculture, National Institute of Food and Agriculture through the ``Artificial Intelligence (AI) Institute for Agriculture” Program under Award AWD003473, and the AgAID Institute. The authors would like to appreciate Dr. Rizwan Qureshi from Center for Research in Computer Vision (CRCV) for his encouraging input.

\section*{Declarations}
The authors declare no conflicts of interest.
\bibliographystyle{elsarticle-harv} 
\bibliography{example}

\vspace{2 cm}

\textbf{Appendix 1}

\onecolumn
{\scriptsize
\tablecaption{Detailed specifications of Large Language Models (2019--2025), including the model name, release year, training data characteristics, and key features.}% 
\begin{longtable}{|c|p{3cm}|p{1.5cm}|c|p{1cm}|p{1cm}|p{1cm}|p{1cm}|p{1cm}|p{1cm}|p{1cm}|}
\caption{\textbf{Comprehensive Architectural Specifications and Transparency Metrics for LLMs: This table presents an in-depth evaluation of language models with a focus on transparency and accessibility. For each model, details include the model name, licensing terms, and weight availability, alongside architectural parameters (layers, hidden units, attention heads, and total parameters). Additionally, performance indicators such as context length, MMLU score, and estimated carbon emissions (tCO\textsubscript{2}eq) are provided.} \label{tab:comparison}}\\
\hline
\textbf{No.} & \textbf{Model} & \textbf{License} & \textbf{Weights} & \textbf{Layers} & \textbf{Hidden} & \textbf{Heads} & \textbf{Params} & \textbf{Context} & \textbf{MMLU score} & \textbf{Carbon Emitted (tCO2eq)} \\
\hline
\endfirsthead
\multicolumn{11}{|c|}{{\bfseries \tablename\ \thetable{} -- continued from previous page}} \\
\hline
\textbf{No.} & \textbf{Model} & \textbf{License} & \textbf{Weights} & \textbf{Layers} & \textbf{Hidden} & \textbf{Heads} & \textbf{Params} & \textbf{Context} & \textbf{MMLU score} & \textbf{Carbon Emitted (tCO2eq)} \\
\hline
\endhead
\hline
\multicolumn{11}{|r|}{{Continued on next column/page}} \\  
\hline
\endfoot
\hline
\endlastfoot
1  & GPT-2 \cite{brown2020language}                              & MIT         & \checkmark   & 48    & 1600   & 25    & 1.5B  & 1024 & N/A     & \ding{55}            \\
 
2  & Legacy ChatGPT-3.5                                         & Proprietary & No           & 96    & 12288  & 96    &  175B & 4K   & 70.0\%  & x (not reported)      \\
 
3  & Default ChatGPT-3.5                                        & Proprietary & No           & 96    & 12288  & 96    &  175B & 4K   & 69.5\%  & 552                  \\
 
4  & GPT-3.5 Turbo                                             & Proprietary & No           & 96    & 12288  & 96    &  175B & 16K  & 71.2\%  & 552                  \\
 
5  & ChatGPT-4                                                 & Proprietary & No           & 96    & 12288  & 96    &  1.8T & 8K   & 86.4\%  & 552                  \\
 
6  & GPT-4o \cite{hurst2024gpt}                                  & Proprietary & No           & 96    & 12288  & 96    &  1.8T & 128K & 88.9\%  & 1,035                \\
 
7  & GPT-4o mini                                               & Proprietary & No           & 96    & 12288  & 96    &  1.2T & 128K & 82.0\%  & \ding{55}            \\
 
8  & o1-preview \cite{jaech2024openai}                           & Proprietary & No           & 128   & 16384  & 128   &  2T   & 128K & 91.3\%  & \ding{55}            \\
 
9  & o1-mini                                                   & Proprietary & No           & 128   & 16384  & 128   &  1.5T & 65K  & 89.5\%  & \ding{55}            \\
 
10 & o1                                                        & Proprietary & No           & 128   & 16384  & 128   &  2.5T & 128K & 92.7\%  & \ding{55}            \\
 
11 & o1 pro mode                                               & Proprietary & No           & 128   & 16384  & 128   &  3T   & 128K & 94.0\%  & \ding{55}            \\
 
12 & o3-mini                                                   & Proprietary & No           & 128   & 16384  & 128   &  1.8T & 128K & 90.1\%  & \ding{55}            \\
 
13 & o3-mini-high                                              & Proprietary & No           & 128   & 16384  & 128   &  1.8T & 128K & 91.5\%  & \ding{55}            \\
 
14 & DeepSeek-R1 \cite{guo2025deepseek}                         & Apache 2.0  & \checkmark   & 64    & 8192   & 64/8  & 671B  & 128K & 90.8\%  & 44                   \\
 
15 & DeepSeek LLM \cite{bi2024deepseek}                          & Proprietary & \checkmark   & 24    & 2048   & 16    & 67B   & 2048 & N/A     & 44                   \\
 
16 & DeepSeek LLM V2 \cite{liu2024deepseek}                      & Proprietary & No           & Not specified & Not specified & Not specified & 236B  & 128K & 78.5\%  & \ding{55}            \\
 
17 & DeepSeek Coder V2 \cite{zhu2024deepseek}                    & Proprietary & No           & Not specified & Not specified & Not specified & 236B  & 128K & 79.2\%  & \ding{55}            \\
 
18 & DeepSeek V3 \cite{liu2024deepseek3}                         & Proprietary & No           & Not specified & Not specified & Not specified & 671B  & 128K & 75.7\%  & \ding{55}            \\
 
19 & BERT-Base \cite{devlin2019bertpretrainingdeepbidirectional}  & Apache 2.0  & \checkmark   & 12    & 768    & 12    & 110M  & 512  & 67.2\%  & 0.652                \\
 
20 & BERT-Large \cite{devlin2019bertpretrainingdeepbidirectional} & Apache 2.0  & \checkmark   & 24    & 1024   & 16    & 340M  & 512  & 69.3\%  & 0.652                \\
 
21 & T5-Small \cite{raffel2020exploring}                          & Apache 2.0  & Yes          & 6/6   & 512    & 8     & 60M   & 512  & \ding{55} & \ding{55}          \\
 
22 & T5-Base \cite{raffel2020exploring}                           & Apache 2.0  & Yes          & 12/12 & 768    & 12    & 220M  & 512  & 35.9\%  & \ding{55}            \\
 
23 & T5-Large \cite{raffel2020exploring}                          & Apache 2.0  & Yes          & 24/24 & 1024   & 16    & 770M  & 512  & 40\%    & \ding{55}            \\
 
24 & T5-3B \cite{raffel2020exploring}                             & Apache 2.0  & Yes          & 24/24 & 1024   & 32    & 3B    & 512  & \ding{55} & \ding{55}          \\
 
25 & T5-11B \cite{raffel2020exploring}                            & Apache 2.0  & Yes          & 24/24 & 1024   & 128   & 11B   & 512  & 48.6\%  & T5-11B               \\
 
26 & Mistral 7B \cite{jiang2023mistral}                           & Apache 2.0  & \checkmark   & 32    & 4096   & 32    & 7.3B  & 8K   & 62.5\%  & \ding{55}            \\
 
27 & LLaMA 2 70B \cite{touvron2023llama}                          & Llama 2     & \checkmark   & 80    & 8192   & 64    & 65.2B & 4K   & 68.9\%  & 291.42               \\
 
28 & CriticGPT \cite{mcaleese2024llm}                             & Proprietary & $\times$   & Not specified & Not specified & Not specified & Not specified & Not specified & \ding{55} & 552                  \\
 
29 & Olympus                                                   & Proprietary & $\times$   & Not specified & Not specified & Not specified & 2000B & Not specified & \ding{55} & \ding{55}           \\
 
30 & HLAT \cite{fan2024hlat}                                     & Proprietary & $\times$   & Not specified & Not specified & Not specified & 7B    & Not specified & \ding{55} & \ding{55}           \\
 
31 & Multimodal-CoT \cite{zhang2023multimodal}                  & Proprietary & $\times$   & Not specified & Not specified & Not specified & Not specified & Not specified & \ding{55} & \ding{55}           \\
 
32 & AlexaTM 20B \cite{soltan2022alexatm}                        & Proprietary & $\times$   & Not specified & Not specified & Not specified & 20B   & Not specified & \ding{55} & \ding{55}           \\
 
33 & Chameleon \cite{team2024chameleon}                           & Proprietary & $\times$   & Not specified & Not specified & Not specified & 34B   & Not specified & \ding{55} & \ding{55}           \\
 
34 & Llama 3 70B \cite{metaIntroducingMeta}                       & Llama 3     & \checkmark   & Not specified & Not specified & Not specified & 70B   & Not specified & 82.0\%  & 1900                 \\
 
35 & LIMA \cite{zhou2024lima}                                     & Proprietary & $\times$   & Not specified & Not specified & Not specified & 65B   & Not specified & \ding{55} & \ding{55}           \\
 
36 & BlenderBot 3x \cite{xu2023improving}                         & Proprietary & $\times$   & Not specified & Not specified & Not specified & 150B  & Not specified & \ding{55} & \ding{55}           \\
 
37 & Atlas \cite{izacard2023atlas}                                & Proprietary & $\times$   & Not specified & Not specified & Not specified & 11B   & Not specified & 47.9\%  & \ding{55}           \\
 
38 & InCoder \cite{fried2022incoder}                              & Proprietary & $\times$   & Not specified & Not specified & Not specified & 6.7B  & Not specified & \ding{55} & \ding{55}           \\
 
39 & 4M-21 \cite{bachmann20244m}                                  & Proprietary & $\times$   & Not specified & Not specified & Not specified & 3B    & Not specified & \ding{55} & \ding{55}           \\
 
40 & Apple On-Device model \cite{mehta2024openelm}               & Proprietary & $\times$   & Not specified & Not specified & Not specified & 3.04B & Not specified & \ding{55} & \ding{55}           \\
 
41 & MM1 \cite{mckinzie2024mm1}                                   & Proprietary & $\times$   & Not specified & Not specified & Not specified & 30B   & Not specified & \ding{55} & \ding{55}           \\
 
42 & ReALM-3B \cite{moniz2024realm}                              & Proprietary & $\times$   & Not specified & Not specified & Not specified & 3B    & Not specified & \ding{55} & \ding{55}           \\
 
43 & Ferret-UI \cite{you2024ferret}                              & Proprietary & $\times$   & Not specified & Not specified & Not specified & 13B   & Not specified & \ding{55} & \ding{55}           \\
 
44 & MGIE \cite{fu2023guiding}                                   & Proprietary & $\times$   & Not specified & Not specified & Not specified & 7B    & Not specified & \ding{55} & \ding{55}           \\
 
45 & Ferret \cite{you2023ferret}                                 & Proprietary & $\times$   & Not specified & Not specified & Not specified & 13B   & Not specified & \ding{55} & \ding{55}           \\
 
46 & Nemotron-4 340B \cite{adler2024nemotron}                     & Proprietary & $\times$   & Not specified & Not specified & Not specified & 340B  & Not specified & \ding{55} & \ding{55}           \\
 
47 & VIMA \cite{jiang2023vima}                                   & Proprietary & $\times$   & Not specified & Not specified & Not specified & 0.2B  & Not specified & \ding{55} & \ding{55}           \\
 
48 & Retro 48B \cite{wang2023instructretro}                     & Proprietary & $\times$   & Not specified & Not specified & Not specified & 48B   & Not specified & \ding{55} & \ding{55}           \\
 
49 & Raven \cite{huang2023raven}                                 & Proprietary & $\times$   & Not specified & Not specified & Not specified & 11B   & Not specified & \ding{55} & \ding{55}           \\
 
50 & Gemini 1.5 \cite{reid2024gemini}                             & Proprietary & $\times$   & Not specified & Not specified & Not specified & Not specified & Not specified & 90\%    & \ding{55}           \\
 
51 & Med-Gemini-L 1.0 \cite{saab2024capabilities}                & Proprietary & $\times$   & Not specified & Not specified & Not specified & 1500B & Not specified & \ding{55} & \ding{55}           \\
 
52 & Hawk \cite{de2024griffin}                                   & Proprietary & $\times$   & Not specified & Not specified & Not specified & 7B    & Not specified & \ding{55} & \ding{55}           \\
 
53 & Griffin \cite{de2024griffin}                                & Proprietary & $\times$   & Not specified & Not specified & Not specified & 14B   & Not specified & \ding{55} & \ding{55}           \\
 
54 & Gemma \cite{team2024gemma}                                  & Proprietary & $\times$   & Not specified & Not specified & Not specified & 7B    & Not specified & 64.3\%  & \ding{55}           \\
 
55 & Gemini 1.5 Pro \cite{reid2024gemini}                        & Proprietary & $\times$   & Not specified & Not specified & Not specified & 1500B & Not specified & \ding{55} & \ding{55}           \\
 
56 & PaLi-3 \cite{chen2023pali}                                  & Proprietary & $\times$   & Not specified & Not specified & Not specified & 6B    & Not specified & \ding{55} & \ding{55}           \\
 
57 & RT-X \cite{padalkar2023open}                                & Proprietary & $\times$   & Not specified & Not specified & Not specified & 55B   & Not specified & \ding{55} & \ding{55}           \\
 
58 & Med-PaLM M \cite{tu2024towards}                             & Proprietary & $\times$   & Not specified & Not specified & Not specified & 540B  & Not specified & \ding{55} & \ding{55}           \\
 
59 & MAI-1 \cite{MAI2024}                                         & Proprietary & $\times$   & Not specified & Not specified & Not specified & 500B  & Not specified & \ding{55} & \ding{55}           \\
 
60 & YOCO \cite{sun2024you}                                       & Proprietary & $\times$   & Not specified & Not specified & Not specified & 3B    & Not specified & \ding{55} & \ding{55}           \\
 
61 & phi-3-medium \cite{abdin2024phi}                             & Proprietary & $\times$   & Not specified & Not specified & Not specified & 14B   & Not specified & \ding{55} & \ding{55}           \\
 
62 & phi-3-mini \cite{abdin2024phi}                               & Proprietary & $\times$   & Not specified & Not specified & Not specified & 3.8B  & Not specified & \ding{55} & \ding{55}           \\
 
63 & WizardLM-2-8x22B \cite{FACE_2015}                            & Proprietary & $\times$   & Not specified & Not specified & Not specified & 141B  & Not specified & \ding{55} & \ding{55}           \\
 
64 & WaveCoder-Pro-6.7B \cite{yu2023wavecoder}                   & Proprietary & $\times$   & Not specified & Not specified & Not specified & 6.7B  & Not specified & \ding{55} & \ding{55}           \\
 
65 & WaveCoder-Ultra-6.7B \cite{yu2023wavecoder}                 & Proprietary & $\times$   & Not specified & Not specified & Not specified & 6.7B  & Not specified & \ding{55} & \ding{55}           \\
 
66 & WaveCoder-SC-15B \cite{yu2023wavecoder}                     & Proprietary & $\times$   & Not specified & Not specified & Not specified & 15B   & Not specified & \ding{55} & \ding{55}           \\
 
67 & OCRA 2 \cite{mitra2023orca}                                  & Proprietary & $\times$   & Not specified & Not specified & Not specified & 7B, 13B & Not specified & \ding{55} & \ding{55}         \\
 
68 & Florence-2 \cite{xiao2024florence}                          & Proprietary & $\times$   & Not specified & Not specified & Not specified & Not specified & Not specified & \ding{55} & \ding{55}         \\
 
69 & Qwen \cite{bai2023qwen}                                     & Proprietary & $\times$   & Not specified & Not specified & Not specified & 72B   & Not specified & \ding{55} & \ding{55}           \\
 
70 & SeaLLM-13b \cite{nguyen2023seallms}                         & Proprietary & $\times$   & Not specified & Not specified & Not specified & 13B   & Not specified & \ding{55} & \ding{55}           \\
 
71 & Grok-1 \cite{xai2024grok1}                                  & Apache 2.0  & \checkmark   & 64    & 6144   & 48/8  & 314B  & 8K   & N/A     & x                    \\
 
72 & Phi-4 \cite{abdin2024phi}                                   & MIT         & \checkmark   & 48    & 3072   & 32    & 14B   & 16K  & 71.2\%  & x                    \\
 
73 & Megatron-LM \cite{shoeybi2019megatron}                      & Custom      & No           & 72    & 3072   & 32    & 8.3B  & 2048 & \ding{55} & \ding{55}           \\
 
74 & Turing-NLG \cite{smith2022using}                            & Proprietary & No           & 78    & 4256   & 28    & 17B   & 1024 & \ding{55} & \ding{55}           \\
 
75 & CTRL(Conditional Transformer Language Model) \cite{keskar2019ctrl} & Apache 2.0  & \checkmark   & 48    & 1280   & 16    & 1.6B  & 256  & \ding{55} & \ding{55}           \\
 
76 & XLNet \cite{yang2019xlnet}                                   & Apache 2.0  & \checkmark   & 24    & 1024   & 16    & 340M (Base), 1.5B (Large) & 512  & \ding{55} & 0.652              \\
 
77 & RoBERTa \cite{liu2019roberta}                               & MIT         & \checkmark   & 24    & 1024   & 16    & 355M  & 512  & \ding{55} & \ding{55}           \\
 
78 & ELECTRA \cite{clark2020electra}                             & Apache 2.0  & \checkmark   & 12 (Base), 24 (Large) & 768 (Base), 1024 (Large) & 12 (Base), 16 (Large) & 110M (Base), 335M (Large) & 512  & \ding{55} & 0.652 \\
 
79 & ALBERT (A Lite BERT) \cite{lan2019albert}                   & Apache 2.0  & \checkmark   & 12 (Base), 24 (Large) & 768 (Base), 1024 (Large) & 12 (Base), 16 (Large) & 12M (Base), 18M (Large)   & 512  & \ding{55} & 0.652 \\
 
80 & DistilBERT \cite{sanh2019distilbert}                        & Apache 2.0  & \checkmark   & 6     & 768    & 12    & 66M   & 512  & \ding{55} & 0.652 \\
 
81 & BigBird \cite{zaheer2020big}                                & Apache 2.0  & \checkmark   & 12 (Base), 24 (Large) & 768 (Base), 1024 (Large) & 12 (Base), 16 (Large) & 110M (Base), 340M (Large) & 4096 & \ding{55} & \ding{55} \\
 
82 & Gopher \cite{rae2021scaling}                                & Proprietary & No           & 80    & 8192   & 128   & 280B  & 2048 & 60\%    & \ding{55}           \\
 
83 & Chinchilla \cite{hoffmann2022training}                      & Proprietary & No           & 80    & 8192   & 128   & 70B   & 2048 & \ding{55} & \ding{55}           \\
 
84 & PaLM \cite{chowdhery2023palm}                                & Proprietary & No           & 118   & 18432  & 128   & 540B  & 8192 & 69.3\%  & \ding{55}           \\
 
85 & OPT (Open Pretrained Transformer) \cite{zhang2022opt}       & Non-commercial & \checkmark & 96    & 12288  & 96    & 175B  & 2048 & \ding{55} & \ding{55}           \\
 
86 & BLOOM \cite{workshop2022bloom}                               & Responsible AI License & \checkmark & 70    & 14336  & 112   & 176B  & 2048 & 90\%    & \ding{55}           \\
 
87 & Jurassic-1 \cite{lieber2021jurassic}                        & Proprietary & No           & 76    & 12288  & 96    & 178B  & 2048 & 67.5   & \ding{55}           \\
 
88 & Codex \cite{chen2021evaluating}                              & Proprietary & No           & 96    & 12288  & 96    & 12B   & 4096 & \ding{55} & \ding{55}           \\
 
89 & T0 (T5 for Zero-Shot Tasks) \cite{sanh2021multitask}         & Apache 2.0  & \checkmark   & 24    & 1024   & 16    & 11B   & 512  & \ding{55} & \ding{55}           \\
 
90 & UL2 (Unifying Language Learning Paradigms) \cite{tay2022ul2}   & Apache 2.0  & \checkmark   & 32    & 4096   & 32    & 20B   & 2048 & \ding{55} & \ding{55}           \\
 
91 & GLaM (Generalist Language Model) \cite{du2022glam}           & Proprietary & No           & 64    & 8192   & 128   & 1.2T (sparse) & 2048 & \ding{55} & \ding{55}           \\
 
92 & ERNIE 3.0 \cite{sun2021ernie}                               & Proprietary & No           & 48    & 4096   & 64    & 10B   & 512  & \ding{55} & \ding{55}           \\
 
93 & GPT-NeoX \cite{black2022gpt}                                & Apache 2.0  & \checkmark   & 44    & 6144   & 64    & 20B   & 2048 & 33.6   & \ding{55}           \\
 
94 & CodeGen \cite{nijkamp2022codegen}                           & Apache 2.0  & \checkmark   & 32    & 4096   & 32    & 16B   & 2048 & \ding{55} & \ding{55}           \\
 
95 & FLAN-T5 \cite{chung2024scaling}                             & Apache 2.0  & \checkmark   & 24    & 1024   & 16    & 11B   & 512  & 52.5   & 552                  \\
 
96 & mT5 (Multilingual T5) \cite{xue2020mt5}                     & Apache 2.0  & \checkmark   & 24    & 1024   & 16    & 13B   & 512  & 52.4   & 552                  \\
 
97 & Reformer \cite{kitaev2020reformer}                          & Apache 2.0  & \checkmark   & 12    & 768    & 12    & 150M  & 64K  & \ding{55} & 552                  \\
 
98 & Longformer \cite{beltagy2020longformer}                      & Apache 2.0  & \checkmark   & 12    & 768    & 12    & 150M  & 4096 & \ding{55} & 552                  \\
 
99 & DeBERTa \cite{he2020deberta}                                & MIT         & \checkmark   & 12    & 768    & 12    & 1.5B  & 512  & \ding{55} & 552                  \\
 
100 & T-NLG (Turing Natural Language Generation) \cite{TURING2020}  & Proprietary & No           & 78    & 4256   & 28    & 17B   & 1024 & \ding{55} & \ding{55}          \\
 
101 & Switch Transformer \cite{fedus2022switch}                  & Apache 2.0  & \checkmark   & 24    & 4096   & 32    & 1.6T (sparse) & 2048 & \ding{55} & \ding{55}          \\
 
102 & WuDao 2.0 \cite{waoDao2021}                                & Proprietary & No           & 128   & 12288  & 96    & 1.75T & 2048 & 86.4\% & \ding{55}          \\
 
103 & LaMDA \cite{thoppilan2022lamda}                             & Proprietary & No           & 64    & 8192   & 128   & 137B  & 2048 & 86\%   & 552                  \\
 
104 & MT-NLG \cite{smith2022using}                               & Proprietary & No           & 105   & 20480  & 128   & 530B  & 2048 & 67.5\% & 284                  \\
 
105 & GShard \cite{lepikhin2020gshard}                           & Proprietary & No           & 64    & 8192   & 128   & 600B  & 2048 & \ding{55} & 4.3\%              \\
 
106 & T5-XXL \cite{raffel2020exploring}                           & Apache 2.0  & \checkmark   & 24    & 1024   & 16    & 11B   & 512  & 48.6\% & \ding{55}            \\
 
107 & ProphetNet \cite{qi2020prophetnet}                          & MIT         & \checkmark   & 12    & 768    & 12    & 300M  & 512  & \ding{55} & \ding{55}          \\
 
108 & DialoGPT \cite{zhang2019dialogpt}                           & MIT         & \checkmark   & 24    & 1024   & 16    & 345M  & 1024 & 25.81\%& 552                  \\
 
109 & BART \cite{lewis2019bart}                                   & MIT         & \checkmark   & 12    & 1024   & 16    & 406M  & 1024 & \ding{55} & \ding{55}          \\
 
110 & PEGASUS \cite{zhang2020pegasus}                             & Apache 2.0  & \checkmark   & 16    & 1024   & 16    & 568M  & 512  & \ding{55} & \ding{55}          \\
 
111 & UniLM \cite{dong2019unified}                                & MIT         & \checkmark   & 12    & 768    & 12    & 340M  & 512  & \ding{55} & \ding{55}          \\

112 & Grok 3 & Proprietary & \ding{55} & Unknown & Unknown & Unknown & Trillions & Unknown & \ding{55} & Unknown
 
\end{longtable}

%% else use the following coding to input the bibitems directly in the
%% TeX file.

%%\begin{thebibliography}{00}

%% \bibitem[Author(year)]{label}
%% For example:

%% \bibitem[Aladro et al.(2015)]{Aladro15} Aladro, R., Martín, S., Riquelme, D., et al. 2015, \aas, 579, A101

%%\end{thebibliography}

\end{document}